\documentclass[structabstract]{aa}  
\usepackage{graphicx}
\usepackage{txfonts}
\usepackage{wrapfig, subfig, threeparttable, rotating}
\usepackage{natbib, lscape} %hyperref
\bibpunct{(}{)}{,}{a}{}{,} %to follow the A&A style 
\begin{document}
   \title{\emph{Herschel}\thanks{\emph{Herschel} is an ESA space observatory with science instruments provided by European-led Principal Investigator consortia and with important participation from NASA.} PACS observations of shocked gas associated with the
 jets of L1448 and L1157}

   \author{G. Santangelo
          \inst{1}
          \and
          B. Nisini
          \inst{1}
          \and
          S. Antoniucci
          \inst{1}
          \and
          C. Codella
          \inst{2}
          \and
          S. Cabrit
          \inst{3}
          \and
          T. Giannini
          \inst{1}
          \and
          G. Herczeg
          \inst{4}
          \and
          R. Liseau
          \inst{5}
          \and
          M. Tafalla
          \inst{6}
          \and
          E.F. van Dishoeck
          \inst{7,8}
          }

   \institute{Osservatorio Astronomico di Roma, via di Frascati 33, 
              00040 Monteporzio Catone, Italy\\
              \email{gina.santangelo@oa-roma.inaf.it}
         \and
              Osservatorio Astrofisico di Arcetri, Largo Enrico Fermi 5, I-50125 Florence, Italy
         \and
              LERMA, Observatoire de Paris, UMR 8112 of the CNRS, 61 Av. de L’Observatoire, 75014 Paris, France
         \and
              Kavli Institute for Astronomy and Astrophysics, Peking University, Yi He Yuan Lu 5, Hai Dian Qu, 100871 Beijing, P.R. China
         \and
              Department of Earth and Space Sciences, Chalmers University of Technology, Onsala Space Observatory, 439 92 Onsala, Sweden
         \and
              Observatorio Astron\'omico Nacional (IGN), Alfonso XII 3, E-28014 Madrid, Spain
         \and
              Leiden Observatory, Leiden University, PO Box 9513, 2300 RA Leiden, The Netherlands
         \and
              Max Planck Institut f{\"u}r Extraterrestrische Physik, Giessenbachstrasse 1, 85748 Garching, Germany
             }

   \date{Received ; accepted }

% \abstract{}{}{}{}{} 
% 5 {} token are mandatory
 
  \abstract
   {}
   {In the framework of the Water In Star-forming regions with Herschel (WISH) key program, 
several H$_2$O ($E_{\rm u} > 190$~K), high-$J$ CO, [O{\sc i}], and OH transitions are mapped with 
\emph{Herschel}-PACS in two shock positions along two prototypical outflows around the low-luminosity 
sources L1448 and L1157.
Previous \emph{Herschel}-HIFI H$_2$O observations ($E_{\rm u}= 53-249$~K) are also used.
The aim is to derive a %consistent and 
complete 
picture of the excitation conditions at the 
selected shock positions.
}
   {We adopted a large velocity gradient analysis (LVG)
to derive the physical parameters of the H$_2$O and CO emitting gas.
Complementary Spitzer mid-IR H$_2$ data were used to derive the H$_2$O abundance.
}
   {Consistent with other studies, at all selected shock spots
a close spatial association between H$_2$O, mid-IR H$_2$, and high-$J$ CO emission is found,
whereas the low-$J$ CO emission traces either entrained ambient gas or a remnant of an older shock.
The excitation analysis, conducted in detail at the L1448-B2 position, 
suggests that a two-component model is needed to reproduce the H$_2$O, CO, and mid-IR H$_2$ lines:
an extended warm component ($T \sim 450$~K) is traced by the H$_2$O emission with $E_{\rm u}= 53-137$~K 
and by the CO lines up to $J=22-21$, and a compact hot component ($T=1100$~K) 
is traced by the H$_2$O emission with $E_{\rm u} > 190$~K and by the higher-$J$ CO transitions.
At L1448-B2 we obtain an H$_2$O abundance $(3-4) \times 10^{-6}$ for the warm component 
and $(0.3-1.3) \times 10^{-5}$ for the hot component and a CO abundance of a few $10^{-5}$ in both components.
In L1448-B2 we also detect OH and blue-shifted [O{\sc i}] emission, spatially coincident with the 
other molecular lines and with [Fe{\sc ii}] emission.
This suggests a dissociative shock for these species, related to the embedded atomic jet. 
On the other hand, a non-dissociative shock at the point of impact of the jet on the cloud 
is responsible for the H$_2$O and CO emission.
The other examined shock positions show an H$_2$O excitation similar to L1448-B2, but 
a slightly higher H$_2$O abundance (a factor of $\sim 4$).
}
    {The two gas components may represent a gas stratification in the post-shock region.
%In particular, 
The extended and low-abundance warm component traces the post-shocked gas that has already cooled down 
to a few hundred Kelvin, whereas the compact and possibly higher-abundance hot component is associated with the gas 
that is currently undergoing a shock episode. 
This hot gas component is more affected by evolutionary effects on the timescales of the outflow propagation, 
which explains the observed H$_2$O abundance variations. %at the different positions.
}


   \maketitle
%
%________________________________________________________________

\section{Introduction}

During the earliest stages of star formation, young stars produce fast collimated jets, 
that collide with the dense parent cloud generating strong interstellar shocks.
These processes strongly modify the chemical composition of the surrounding gas 
and are identified by intense line emission.
Among the different tracers of shocks, water is a key molecule and a unique diagnostic tool of 
local conditions and energetic processes occurring in star-forming regions 
\citep[e.g.][]{vandishoeck2011}, since its abundance varies by many orders of magnitude during the shock lifetime 
\citep[e.g.][]{bergin1998}.
In particular, water abundance with respect to H$_2$ is expected to increase from $< 10^{-7}$ 
in cold regions to about $10^{-4}$ in the warm gas, due to the combined effects of evaporation 
of icy mantles and high-temperature chemical reactions which drive all the atomic oxygen into H$_2$O.

Space instruments, such as SWAS, Odin, and ISO, have allowed the study of water in outflows.
It was possible to resolve the water line profiles 
\citep[e.g.][]{benedettini2002,bjerkeli2009}, 
to derive the excitation conditions of the emitting gas 
\citep[e.g.][]{liseau1996,ceccarelli1998,nisini1999,nisini2000} 
and to measure the water abundance in shocks 
through comparison with CO emission. 
In particular, values within the range $\sim 10^{-7}-10^{-4}$ 
have been derived for the H$_2$O abundance showing that it 
depends on the gas temperature and velocity \citep[e.g.][]{giannini2001,franklin2008}.
However, the limited spatial and spectral resolution of these instruments have prevented 
a clear association of the shocked gas with a specific kinematical component and with a 
specific region along the outflow, thus preventing the origin of the shocked gas from being derived.
Thanks to the {\it Herschel Space Observatory}, we are now able to improve our view 
of the shock processes occurring during the very early stages of star formation and 
to test the model predictions for the water formation and abundance during these processes.

In this context, the low-luminosity Class~0 protostellar systems L1448 (7.5~L$_{\odot}$)
and L1157 (4~L$_{\odot}$) are excellent targets.
At the distance of 232~pc \citep{hirota2011}, L1448 is the prototype of a source driving a molecular jet 
\citep{eisloffel2000}. 
It has a powerful and highly collimated outflow, which has been extensively studied 
\citep[e.g.][]{guilloteau1992,bachiller1995,hirano2010}. Gas excited in shocks 
has been detected along the outflow through near- and mid-IR emission of molecular 
hydrogen \citep[e.g.][]{neufeld2009,giannini2011}. 
At a distance of 250~pc, L1157 is perhaps the most active outflow from a chemical point of view 
\citep{bachiller1997,bachiller2001}, often quoted as being the prototype of the so-called chemically 
rich outflows. 
Detailed \emph{Herschel} observations of the L1157 outflow by the Chemical HErschel Surveys of Star forming regions
(CHESS) program have been presented by 
\citet{codella2010,lefloch2010,codella2012a,codella2012b,benedettini2012,lefloch2012}.

The \emph{Herschel} Key Program Water In Star-forming regions with Herschel  
\citep[WISH,][]{vandishoeck2011} employed more than 400~hours of telescope 
time to observe H$_2$O and related molecules toward about 80 protostars at different evolutionary stages and masses 
to study the physical and chemical conditions of the gas in nearby star-forming regions.
Within the WISH framework, several results concerning outflows have been presented 
\citep[e.g.][]{bjerkeli2011,kristensen2011,kristensen2012,bjerkeli2012,herczeg2012,tafalla2013}.
Both the L1448 and L1157 outflows have been mapped to study the spatial 
distribution of water and the results have been presented by \citet{nisini2010a} for 
the L1157 outflow and \citet{nisini2013} for the L1448 outflow.
They show a clumpy water distribution, with emission peaks corresponding to shock positions 
along the outflow. 
Multi-transition observations (with excitation energies ranging from 53 to 249~K), 
performed with the Heterodyne Instrument for the Far Infrared \citep[HIFI,][]{degraauw2010}
toward two shock positions of each outflow, have 
been presented by \citet{vasta2012} for the L1157 outflow and by \citet{santangelo2012} 
for the L1448 outflow to constrain the water excitation conditions.
These studies have shown strong variations of the H$_2$O line profiles with excitation, 
which indicate that gas components with different physical and excitation conditions coexist
at the shock positions.
Complex line profiles have also been observed at the position of the 
central driving source of the L1448 outflow by \citet{kristensen2012}, with a broad velocity component
possibly associated with the interaction of the outflow with the protostellar envelope
and the extreme high-velocity gas (EHV) associated with the collimated molecular jet.

In this context as part of the WISH key program, we report here on the results of 
new \emph{Herschel} observations of the same shock regions along the L1448 and L1157 outflows.
A set of high excitation H$_2$O lines and several transitions of CO, OH, and [O{\sc i}] 
have been mapped with the Photodetecting Array Camera and Spectrometer \citep[PACS,][]{poglitsch2010} instrument. 
Unlike the previous HIFI observations, the PACS data will allow us to detect and 
characterize the higher excitation gas with a higher angular resolution, 
thus providing a complete and consistent picture of the shocked gas along the two outflows.
This in turn will allow us to settle the conditions 
for water formation and to explore its ability to probe specific excitation regimes.

The paper is organized as follows. The PACS observations are described in Sect.~\ref{sect:observations}.
In Sect.~\ref{sect:results} we present the PACS maps and the main observational results. 
A detailed analysis of the PACS maps is discussed in Sect.~\ref{sect:analysis}, 
starting from the study of the physical and excitation 
conditions in the B2 shocked position along the L1448 outflow and subsequently discussing 
the implications for the other selected shocked spots. 
The results are discussed in Sect.~\ref{sect:shock_models}, in the context of current shock models.
Finally, the conclusions are presented in Sect.~\ref{sect:conclusions}.

%
%__________________________________________________________________

\section{Observations and data reduction}
\label{sect:observations}

\begin{figure*}%[ht]
   \centering
   \includegraphics[width=0.90\textwidth]{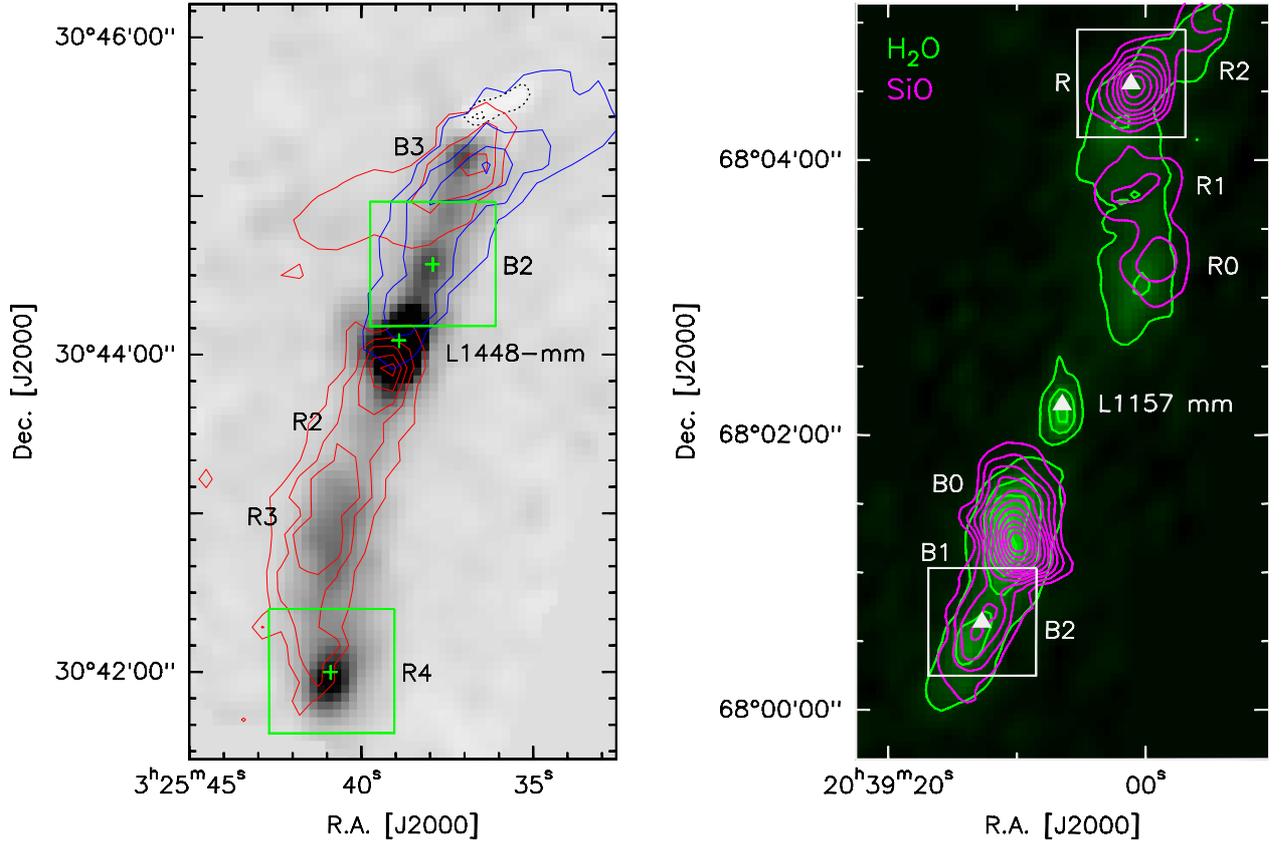}
   \caption{PACS H$_2$O 179~$\mu$m images of L1448 and L1157 \citep{nisini2010a,nisini2013}.
The PACS line survey positions are indicated for 
L1448-B2 and R4 with crosses, for L1157-B2 and R with triangles. 
The field of view of the PACS observations is displayed as a box at the selected positions. 
CO($3-2$) and SiO($3-2$) emissions for L1448 and L1157, respectively, are superimposed on the H$_2$O maps.
}
   \label{fig:outflows}
   \end{figure*}

We performed a survey of key far-IR lines with the PACS instrument on board 
\emph{Herschel} \citep{pilbratt2010,poglitsch2010}
toward two shock spots along each interested outflow (see Fig.~\ref{fig:outflows}): the B2 and R4 spots 
along L1448 \citep[hereafter L1448-B2 and L1448-R4, respectively;][]{bachiller1990};
and B2 and R along L1157 \citep[hereafter L1157-B2 and L1157-R, respectively;][]{bachiller2001}. 
The PACS instrument is an integral field unit (IFU), consisting of a $5 \times 5$ array of spatial pixels (hereafter spaxels). 
Each spaxel covers 9$.\!\!^{\prime\prime}4 \times 9.\!\!^{\prime\prime}4$, providing a total field of view of
$47^{\prime\prime} \times 47^{\prime\prime}$.
The line spectroscopy mode was used to cover short spectral regions 
and thus observe selected lines at the interested shock positions.
The line survey comprises ortho- and para-H$_2$O transitions with excitation energies ranging from 
194 to 396~K. 
In addition, high-$J$ CO, [O{\sc i}], and OH lines have been observed (see Table~\ref{table:intensities} 
for a summary of the targeted lines).
These observations are complementary to 
observations of lower excitation H$_2$O transitions \citep{santangelo2012,vasta2012} 
conducted with the HIFI heterodyne instrument \citep{degraauw2010}
with excitation energies ranging from 53 to 249~K,
and to PACS maps of the H$_2$O $2_{12}-1_{01}$ line at 179.5~$\mu$m along 
the two outflows \citep[][see also Fig.~\ref{fig:outflows}]{nisini2010a,nisini2013}.

The data were processed with the ESA-supported package HIPE\footnote{HIPE is a joint development 
by the \emph{Herschel} Science Ground Segment Consortium, consisting of ESA, the NASA \emph{Herschel} 
Science Center, and the HIFI, PACS, and SPIRE consortia.} 
\citep[\emph{Herschel} Interactive Processing Environment,][]{ott2010} version 4.2 
(except for the data relative to the L1157-R position that were processed with HIPE version 5)
\footnote{We have checked the data processed with the latest version of HIPE (version 10) 
and we find an agreement in the flux densities within $15-20$\%, which is definitely within the PACS calibration uncertainty (30\%).}. 
The observed fluxes were normalized to the telescope background and then converted into 
absolute fluxes using Neptune as a calibrator.
The flux calibration uncertainty of the PACS observations is 30\%, 
based on the flux repeatability for multiple observations of the same target 
in different programs and on cross-calibration with HIFI and ISO. 
Further data reduction, to obtain continuum subtracted line maps,
and the analysis of the data were
performed using IDL and the GILDAS\footnote{http://www.iram.fr/IRAMFR/GILDAS/} software.

The \emph{Herschel} diffraction limit at 179~$\mu$m is 12$.\!\!^{\prime\prime}$6 and for wavelengths 
below 133~$\mu$m it is smaller than the PACS spaxel size of 9$.\!\!^{\prime\prime}$4.
To correct for the different beam sizes in the excitation analysis presented in Sect.~\ref{sect:analysis},
we convolved all maps to the resolution of the transition 
with the longest wavelength, that is 12$.\!\!^{\prime\prime}$6 at 179~$\mu$m, 
and then extracted the fluxes at each selected shock spot.

%
%__________________________________________________________________

\section{Results}
\label{sect:results}

The PACS spectra of all the lines detected in the four examined shocked positions 
are presented in Appendix~\ref{appendix:pacs_maps}, whereas a summary of the main line parameters 
is given in Table~\ref{table:intensities}, along with the fluxes of the detected lines. 
The source L1448-B2 represents the position in which we detected 
the largest number of lines and it is the only position where we detected the OH fundamental line at 119~$\mu$m.

\begin{table*}
\caption{Fluxes of the lines observed with PACS and relative 1~$\sigma$ errors in parentheses. } 
\label{table:intensities}
\centering
\begin{tabular}{l r r r c c c c }
\hline\hline
&&&& L1448 - B2 & L1448 - R4 & L1157 - B2 & L1157 - R\tablefootmark{a} \\
\cline{5-8}
Transition & Frequency & Wavelength & $E_{\rm u}/k_{\rm B}$ & \multicolumn{4}{c}{Flux} \\
& (GHz) & ($\mu$m) & (K) & \multicolumn{4}{c}{(10$^{-15}$~erg s$^{-1}$ cm$^{-2}$)} \\
\hline
[O{\sc i}] $^3P_1 - ^3P_2$      & 4744.78 &  63.2 &  227.7 &    167 (8)     & 27 (6)  & 26 (4)  & 32 (4)  \\
o-H$_2$O $2_{21}-1_{10}$         & 2773.98 & 108.1 &  194.1 &     22 (2)     & 20 (4)  & $<$16   &  9 (2)  \\
CO $24-23$                     & 2756.39 & 108.8 & 1656.5 &     14 (4)     & $<$19   & $<$17   & $<$19   \\
CO $22-21$                     & 2528.17 & 118.6 & 1397.4 &     13 (2)     &  7 (2)  & $<$11   & $<$11   \\
OH $^2\Pi_{3/2} J=5/2^--3/2^+$   & 2514.31 & 119.2 &  120.7 &     17\tablefootmark{b} (2) & $<$11  & $<$11  & $<$11  \\ 
OH $^2\Pi_{3/2} J=5/2^+-3/2^-$   & 2509.95 & 119.4 &  120.5 &       --       &  --     &  --     &  --     \\ 
p-H$_2$O $4_{04}-3_{13}$         & 2391.57 & 125.4 &  319.5 &      6 (2)     &  6 (1)  & $<$11   &  5 (1)  \\
p-H$_2$O $3_{13}-2_{02}$         & 2164.13 & 138.5 &  204.7 &     16 (2)     & 20 (2)  &  5 (1)  & 15 (1)  \\
CO $18-17$                     & 2070.62 & 144.8 &  945.0 &     32 (2)     & 11 (2)  & $<$9    & $<$8    \\
$[$O{\sc i}$]$ $^3P_0 - ^3P_1$  & 2060.07 & 145.5 &  326.6 &      8 (1)     & $<$8    & $<$9    &  6 (3)  \\
CO $16-15$                     & 1841.35 & 162.8 &  751.7 &     39 (2)     & 17 (2)  &  5 (2)   & 10 (2)  \\
o-H$_2$O $3_{03}-2_{12}$         & 1716.77 & 174.6 &  196.8 &     51 (2)     &  77 (1)  & 22 (1)  & 45 (1)  \\
o-H$_2$O $2_{12}-1_{01}$\tablefootmark{c}& 1669.90 & 179.5 & 114.4 & 90 (9)  & 139 (35) & 85 (8)  & 53 (13)  \\
\hline
\end{tabular}
\tablefoot{Fluxes are measured at the central spaxel of the maps after convolving to 12$.\!\!^{\prime\prime}$6,
i.e. the PACS resolution of the transition with the longest wavelength (179~$\mu$m).
The relative rms error is measured at the same spaxel and does not include 30\% calibration accuracy.
\tablefoottext{a}{The fluxes and relative rms errors are given at the peak of the H$_2$O emission, which is at a position 
offset of (2$^{\prime\prime}$,$-9^{\prime\prime}$) from the central spaxel (see Fig.~\ref{fig:l1157mappe}). 
}
\tablefoottext{b}{The value represents the sum of the fluxes of the two listed OH lines (at 119.2 and 119.4~$\mu$m).
The OH 119.4/119.2 line ratio measured at the central spaxel of the map is 1.3.}
\tablefoottext{c}{The values are measured from the PACS maps of the two outflows at 179~$\mu$m \citep{nisini2013}.
}}
\end{table*}

   \begin{figure*}%[ht]
   \centering
   \includegraphics[angle=-90,width=0.83\textwidth]{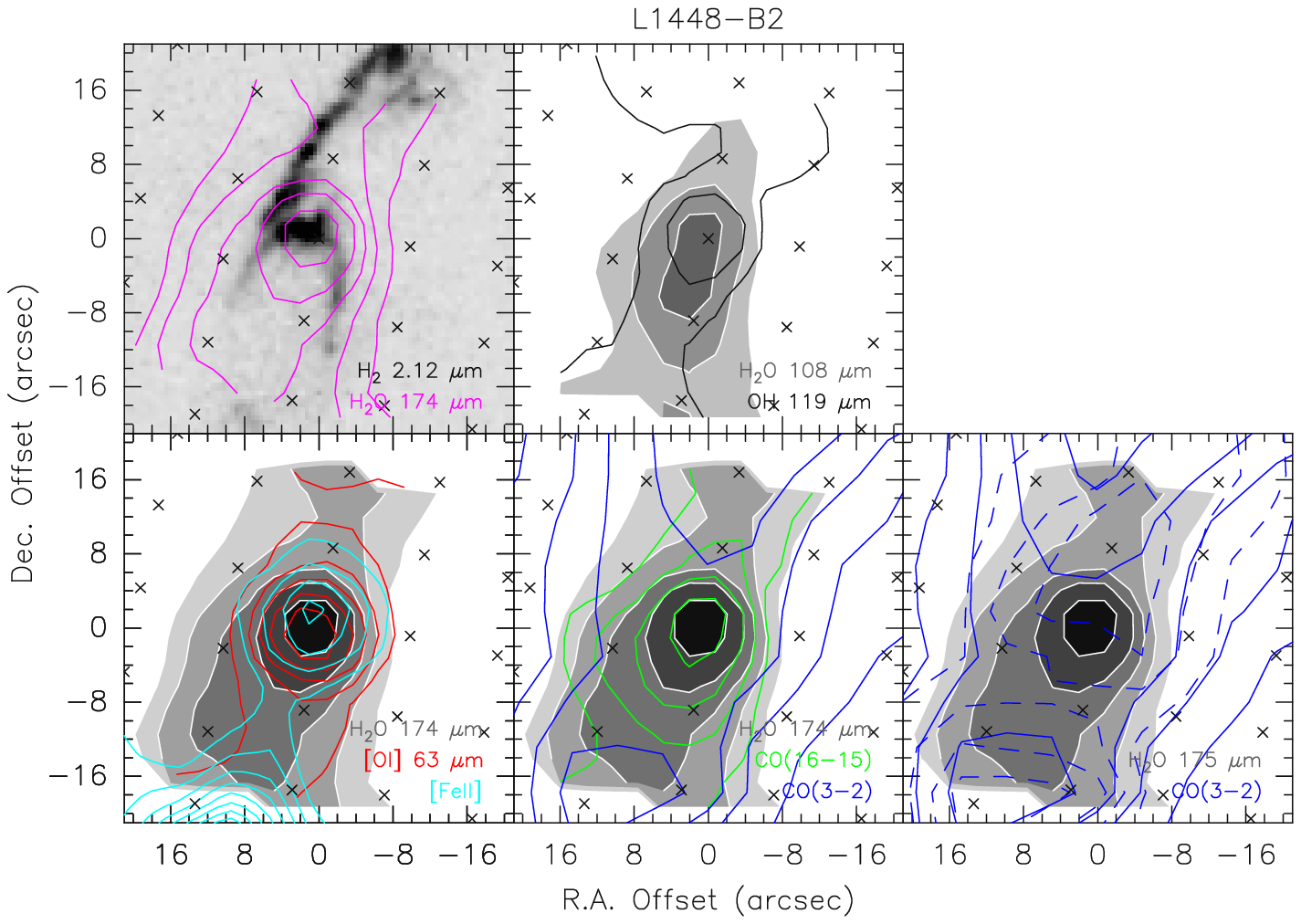}
   \includegraphics[angle=-90,width=0.55\textwidth]{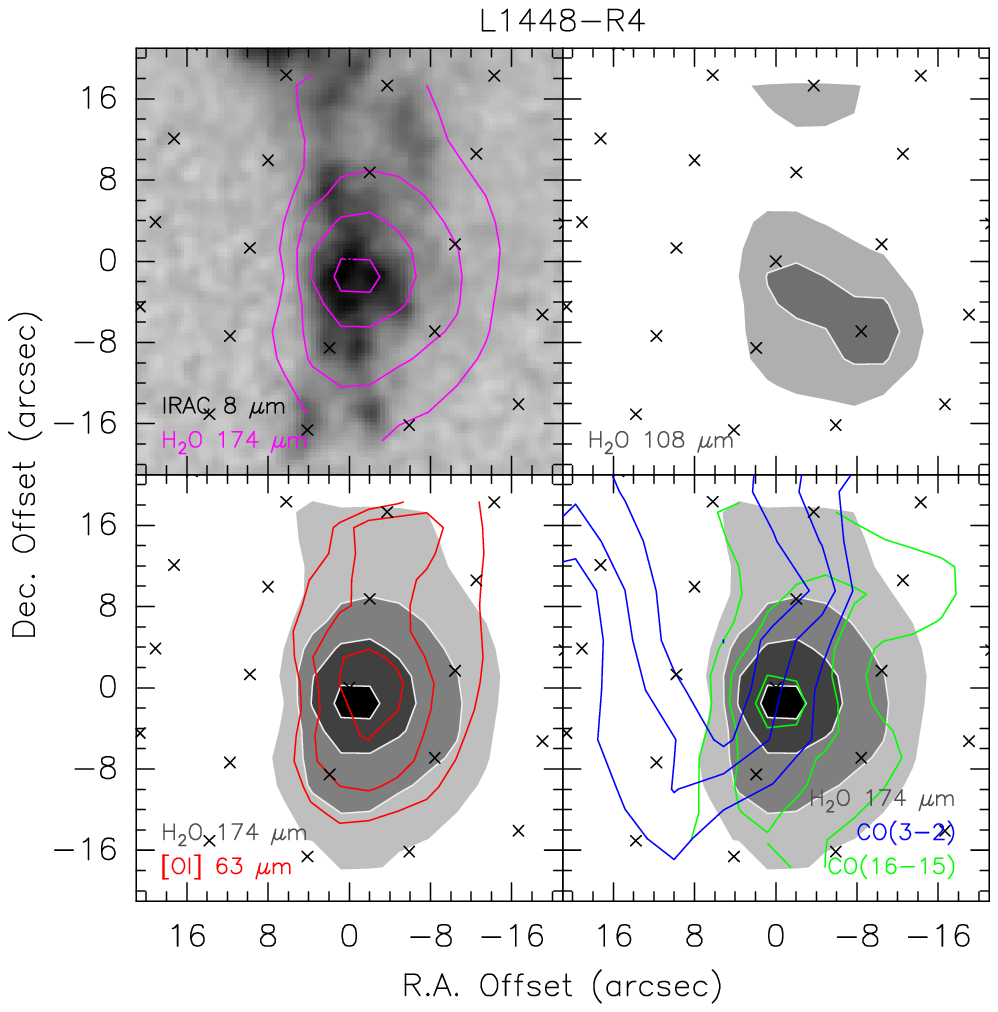}
   \caption{Overlay between PACS H$_2$O $3_{03}-2_{12}$ (174~$\mu$m), H$_2$O $2_{21}-1_{10}$ (108.1~$\mu$m), 
[O{\sc i}] $^3P_1 - ^3P_2$ (63.2~$\mu$m), CO($16-15$), OH (119~$\mu$m),
and other tracers in the B2 (\emph{upper panel}) and R4 (\emph{lower panel}) shocked spots along the L1448 outflow. 
In particular, JCMT CO($3-2$) emission (Half Power Beam Width, HPBW $\sim 14^{\prime\prime}$) from \citet{nisini2013}, 
Spitzer [Fe{\sc ii}] emission at 26~$\mu$m from \citet{neufeld2009}, 
IRAC 8~$\mu$m emission from \citet{tobin2007}, and H$_2$ emission at 2.12~$\mu$m from \citet{davis1995} are shown. 
In the bottom-right panel relative to L1448-B2 the EHV 
CO($3-2$) emission ($v \gtrsim -50$~km~s$^{-1}$) and the HV CO($3-2$) emission 
($v \lesssim -40$~km~s$^{-1}$) are shown in dashed and solid lines, respectively.
The contours in each map are traced every 3~$\sigma$, starting from a 5~$\sigma$ level, 
except for the CO($3-2$) in L1448-B2, where the contours are traced in steps of 5~$\sigma$, starting from 5~$\sigma$.
The crosses represent the pointing of the 25 spaxels.
}
   \label{fig:l1448mappe}
   \end{figure*}
   \begin{figure*}%[ht]
   \centering
   \includegraphics[angle=-90,width=0.6\textwidth]{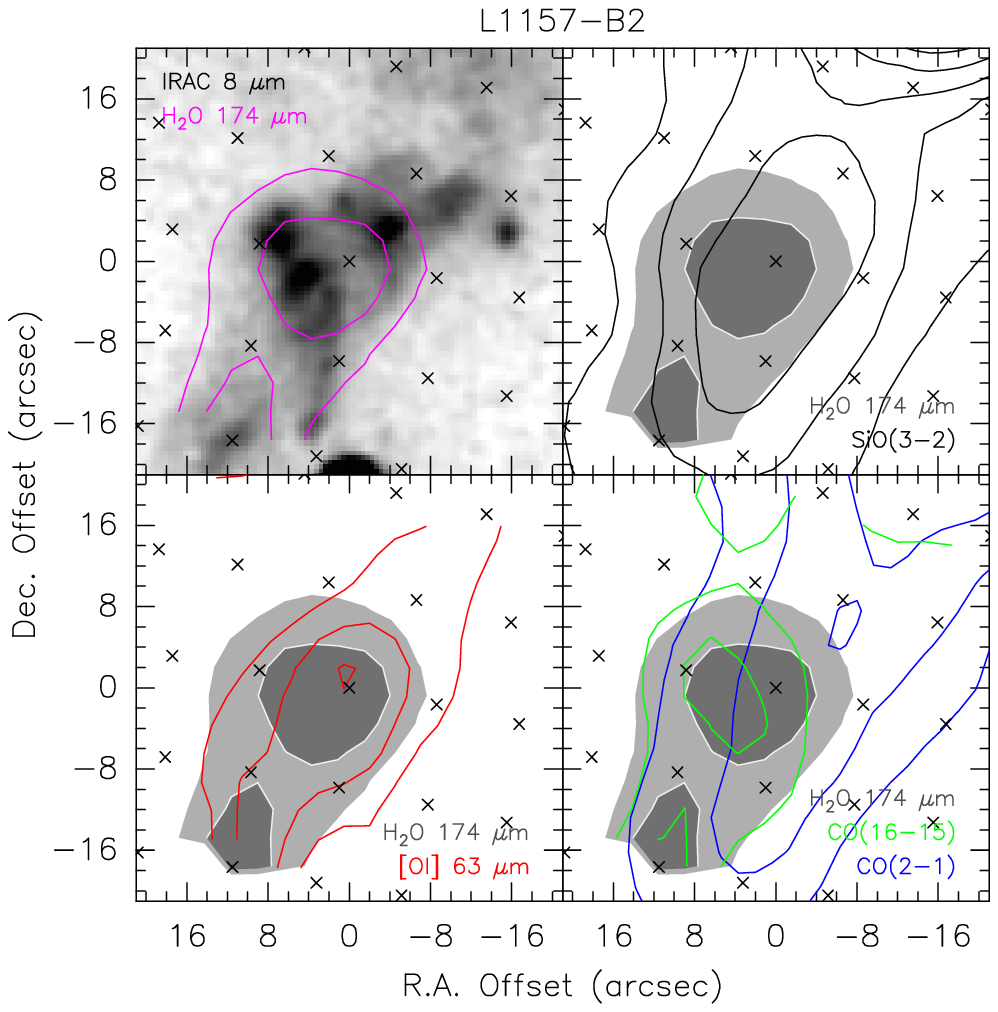}
   \includegraphics[angle=-90,width=0.6\textwidth]{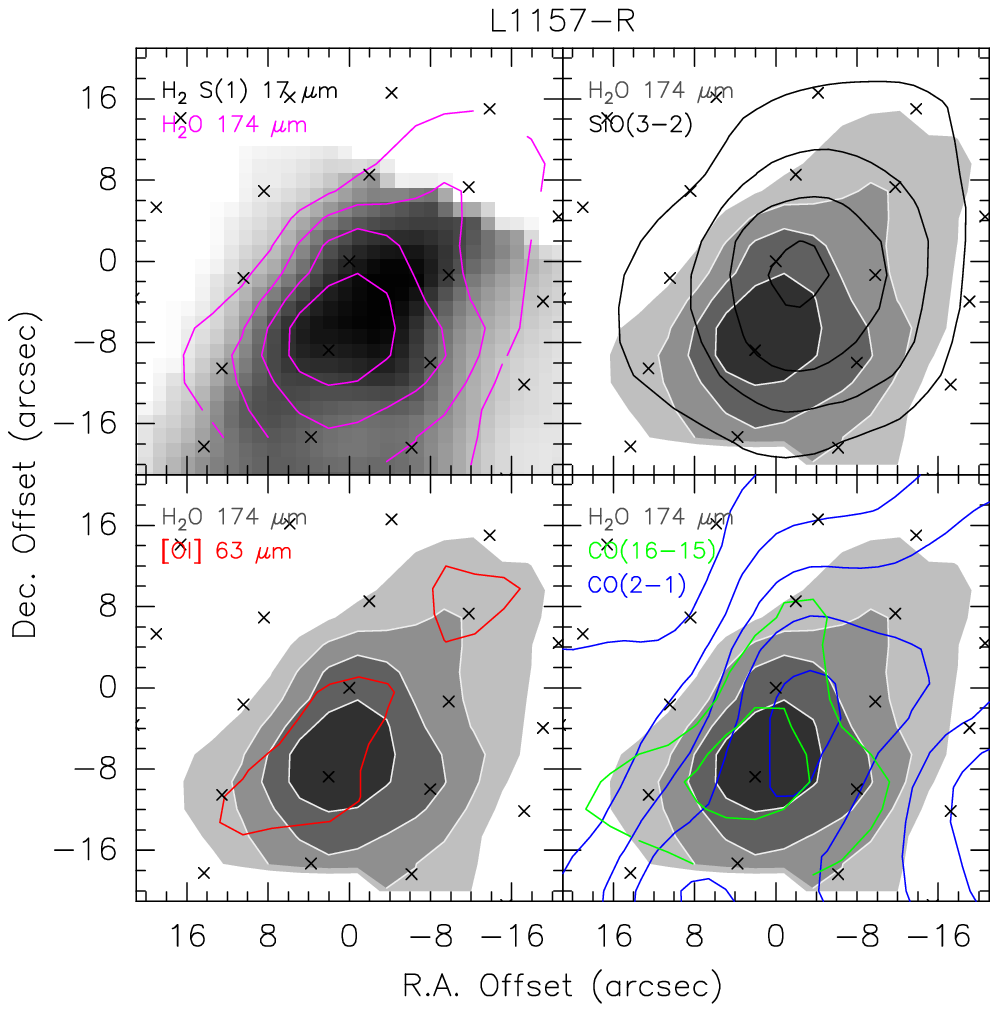}
   \caption{The same as in Fig.~\ref{fig:l1448mappe}, but for the L1157 outflow. 
Contours are displayed for IRAM-30m CO($2-1$) and SiO($3-2$) emission 
(HPBW equal to 11$^{\prime\prime}$ and 18$^{\prime\prime}$, respectively) from \cite{bachiller2001} 
and are traced in steps of 5~$\sigma$, starting from 5~$\sigma$.
Spitzer-IRAC 8~$\mu$m emission and Spitzer-IRS H$_2$ S(1) emission at 17~$\mu$m from \citet{neufeld2009} are shown. 
}
   \label{fig:l1157mappe}
   \end{figure*}

The original PACS maps of selected lines, not convolved to a common angular resolution, 
are shown in Fig.~\ref{fig:l1448mappe} and \ref{fig:l1157mappe}, respectively, for L1448 and L1157.
The figures present the overlay between the H$_2$O $3_{03}-2_{12}$ (174~$\mu$m), H$_2$O $2_{21}-1_{10}$ 
(108.1~$\mu$m),
CO($16-15$), [O{\sc i}] $^3P_1 - ^3P_2$ (63.2~$\mu$m), OH (119~$\mu$m) emission, 
and other tracers from complementary observations.
Along the L1448 outflow several interesting features can be noticed. In particular, 
the peak of the H$_2$O emission at L1448-B2 is at the apex of the bow-shock, as traced by the H$_2$ 
emission at 2.12~$\mu$m, and at L1448-R4 it corresponds to the peak of the IRAC 8~$\mu$m emission.
We found no shift at this angular resolution between the H$_2$O emission at 174~$\mu$m and CO($16-15$) 
emission, both in L1448-B2 and L1448-R4, which indicates that at this angular resolution 
high-$J$ CO and H$_2$O are spatially coincident and trace shocked gas.

The bottom-right panel relative to L1448-B2 shows the comparison between H$_2$O and CO($3-2$). For CO,
the EHV gas \citep[i.e. $v \gtrsim -50$ km~s$^{-1}$,][]{bachiller1990} 
has been separated from the standard outflow high-velocity (HV, $v \lesssim -40$ km~s$^{-1}$) gas emission. 
We see that the HV gas is totally uncorrelated with the water emission, a result already found in other studies
\citep[][]{santangelo2012,nisini2013,tafalla2013}. The EHV gas, on the
other hand, has a peak shifted north-west with respect to the H$_2$O peak. Thus, the low-$J$ 
CO emission traces entrained ambient gas and not the shocked gas, independent of the
velocity components.

Finally, we do not see any significant spatial shift at this angular resolution 
between the peaks of H$_2$O, [O{\sc i}], and [Fe{\sc ii}] emission, although the H$_2$O emission 
appears to be more extended than [O{\sc i}] and [Fe{\sc ii}].
Nevertheless, at the L1448-B2 position, and possibly at the adjacent 
spaxels along the outflow direction, a hint of a velocity shift 
in the [O{\sc i}] line at 63~$\mu$m was detected: 
the line is blue-shifted by $\sim 80$~km~s$^{-1}$ (see Fig.~\ref{fig:PACSl1448B2}), 
which is comparable with the resolution 
element of PACS at this wavelength ($\sim 90$~km~s$^{-1}$).
Similar [O{\sc i}] velocity shifts have been found in HH46 by \citet{vankempen2010a}
and in Serpens SMM1 by \citet{goicoechea2012}, who suggested 
the presence of fast dissociative shocks close to the protostar,
related with an embedded atomic jet.
Moreover, \citet{karska2013} analysed PACS spectra of a large sample of Class 0/I protostars
and they found such profile shifts toward at least 1/3 of their targets.
We point out that Fe has a ionization potential of 7.9~$eV$ and thus 
we expected to find [Fe{\sc ii}] co-spatial with [O{\sc i}]
(ionization potential of 13.6~$eV$).

A much smaller number of lines was detected along the L1157 outflow. 
In particular, only four lines were detected at L1157-B2.
Here the emission is elongated in the outflow direction, according to all tracers.
Similarly to L1448, the H$_2$O emission at 174~$\mu$m is spatially associated 
with the [O{\sc i}] $^3P_1 - ^3P_2$ emission at 63.2~$\mu$m and the CO($16-15$) emission.
Two emission peaks can be identified in the PACS maps: the brightest one 
is found at the central spaxel and is spatially associated with the H$_2$ emission, 
as seen from the overlay with the Spitzer-IRAC image at 8~$\mu$m;
the other emission peak is at a position offset of (12$^{\prime\prime}$,$-18^{\prime\prime}$) from the central spaxel, 
close to the edge of the PACS map.
The SiO($3-2$) emission \citep{bachiller2001} also appears to be elongated along the outflow direction 
with a peak roughly corresponding to the central spaxel of the PACS maps. 
On the other hand, the CO($2-1$) emission is not spatially associated with any other molecular species.
At L1157-R a bright emission peak is seen in H$_2$O 
and in all species observed with PACS.
This H$_2$O peak is shifted with respect to the central spaxel of (2$^{\prime\prime}$,$-9^{\prime\prime}$) and 
is spatially associated with the H$_2$ emission. 
A second peak is found in the [O{\sc i}] $^3P_1 - ^3P_2$ emission (63.2~$\mu$m), at a 
position offset of ($-12^{\prime\prime}$,8$^{\prime\prime}$) from the central spaxel, and 
is also visible in H$_2$O. 
On the other hand, the SiO($3-2$) emission peaks at the central spaxel position, 
thus offset from the H$_2$O emission.
Finally, the CO($2-1$) emission is more diffuse than the other tracers and has an emission peak 
at the central spaxel position, thus shifted with respect to the H$_2$O and high-$J$ CO emission.

In conclusion, the inspection of the PACS maps highlights the following results: 
in both outflows the H$_2$O emission is spatially associated with mid-IR H$_2$ emission 
and high-$J$ CO emission, whereas the low-$J$ CO emission seems to be associated with a 
different gas component.
Our findings are consistent with the results obtained by 
\citet{nisini2010a,nisini2013} from mapping the H$_2$O $2_{12}-1_{01}$ emission along the L1448 and L1157 outflows 
and by \citet{tafalla2013} from the analysis of H$_2$O $1_{10}-1_{01}$ and $2_{12}-1_{01}$ emission in
a large sample of shocked positions.
Moreover, \citet{karska2013} found a tight correlation between H$_2$O 2$_{12}-1_{01}$ at 179~$\mu$m 
and high-$J$ CO line fluxes, concluding that they likely arise in the same gas component.
The SiO emission appears to be slightly shifted with respect to H$_2$O, consistent with the two
gas components tracing shock regions with different excitation conditions, as discussed in \citet{santangelo2012} 
and \citet{vasta2012}.
Finally, no shift is observed at the PACS angular resolution between the [O{\sc i}] and [Fe{\sc ii}] lines 
and the H$_2$O emission.

   \begin{figure}%[ht]
   \centering
   \includegraphics[width=0.45\textwidth]{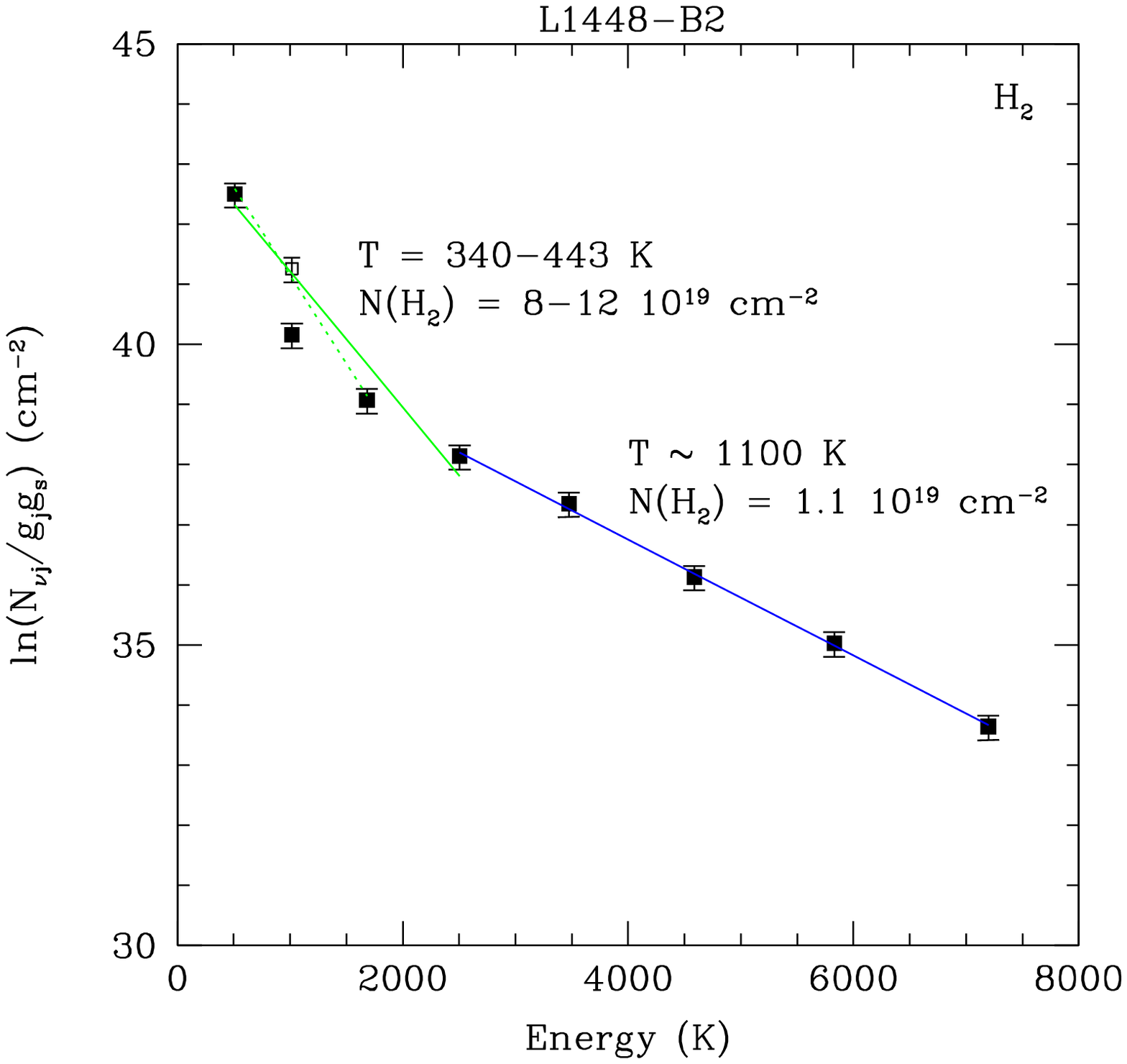}
   \includegraphics[width=0.45\textwidth]{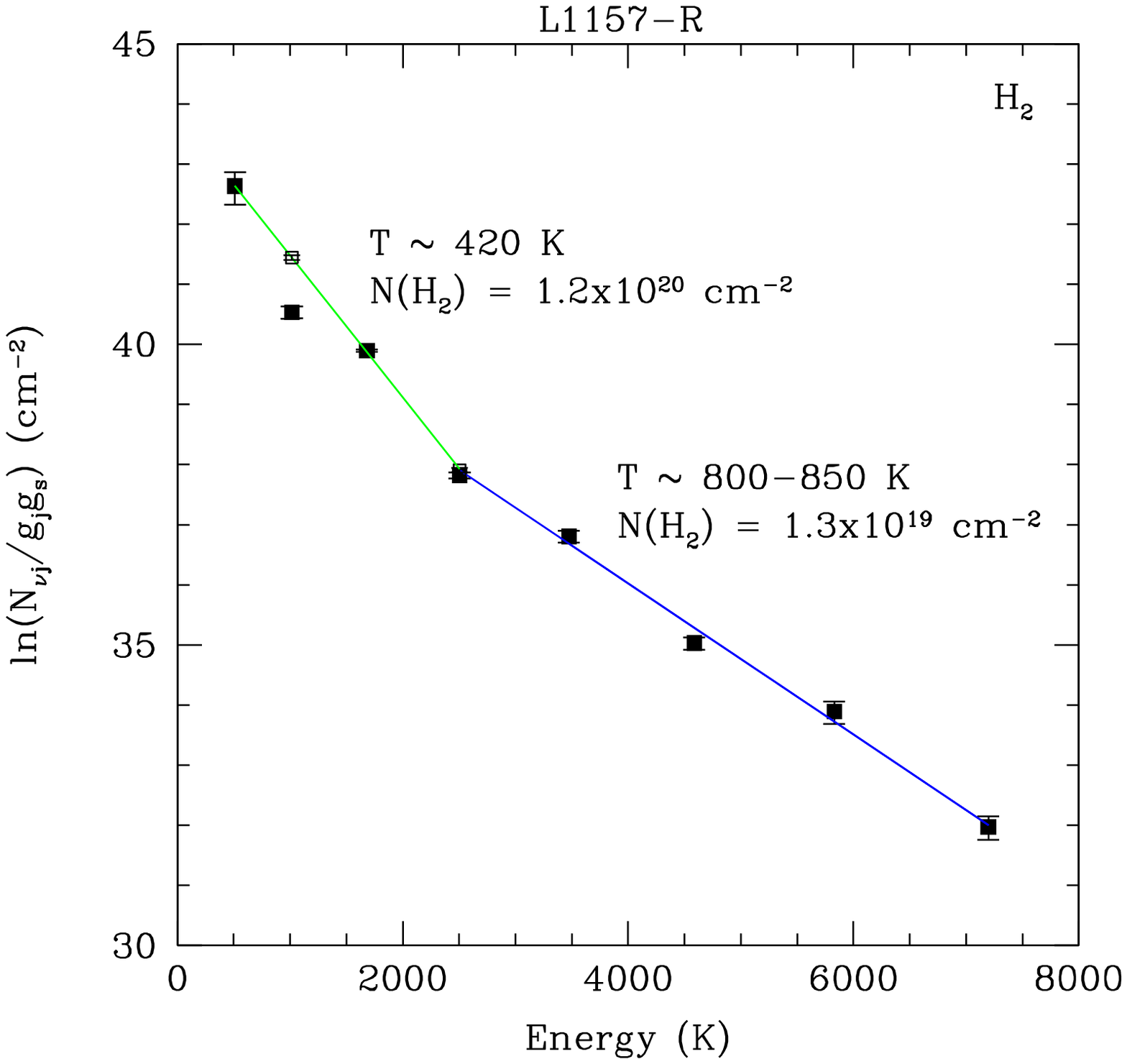}
   \caption{\emph{Upper:} Rotational diagram at L1448-B2 
for the H$_2$ emission lines detected with Spitzer by \citet{giannini2011}.
The values have been derived from the H$_2$ fluxes integrated over a 13$^{\prime\prime}$ area
for comparison with the PACS data.
The black dots are the observed values, whereas the empty dot represents the H$_2$ S(1) line 
corrected for an ortho-to-para ratio equal to 1 \citep[see][]{giannini2011}.
The green solid line represents the linear fit to the S(0)--S(3) H$_2$ lines, 
while the green dotted line is the fit obtained using the S(0)--S(2) H$_2$ lines.
Finally, the blue line is the linear fit to the S(3)--S(7) lines.
The resulting parameters ($T$,$N$) or range of parameters of the linear fits are reported in the diagram.
\emph{Lower:} H$_2$ rotational diagram at L1157-R. The symbols are the same as in the upper panel.
}
   \label{fig:H2_rot-diagr}
   \end{figure}

   \begin{figure}%[ht]
   \centering
   \includegraphics[width=0.39\textwidth]{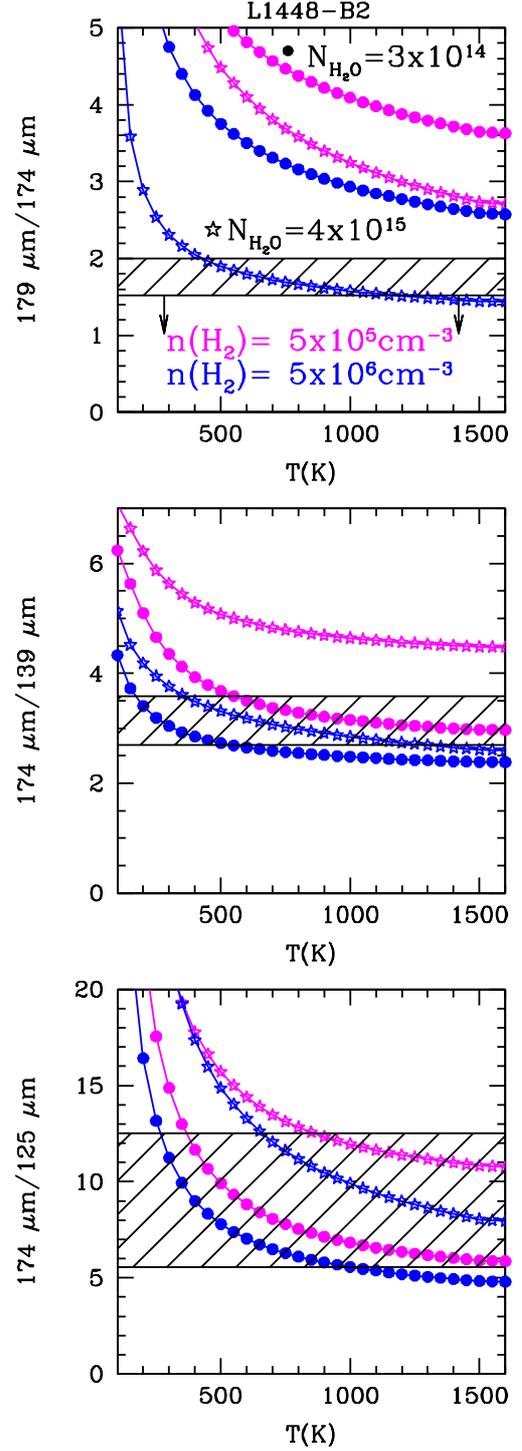}
   \caption{\emph{Top}: Ratio between H$_2$O $2_{12}-1_{01}$ (179~$\mu$m)
and $3_{03}-2_{12}$ (174~$\mu$m) emission, as a function of $T$ for two values of H$_2$ density 
($n_{\rm H_2}=5 \times 10^5$~cm$^{-3}$ and $5 \times 10^6$~cm$^{-3}$) and 
two values of H$_2$O column density ($N_{\rm H_2O}=3 \times 10^{14}$ and 
$4 \times 10^{15}$~cm$^{-2}$) of L1448-B2.
The shaded band highlights the H$_2$O ratio observed with PACS and the arrows indicate that this ratio 
can be considered an upper limit (see text for details).
\emph{Middle} and \emph{Bottom}: Ratios between H$_2$O $3_{03}-2_{12}$ (174~$\mu$m) and $3_{13}-2_{02}$ (138.5~$\mu$m)
emission and between H$_2$O $3_{03}-2_{12}$ (174~$\mu$m) and $4_{04}-3_{13}$ (125.4~$\mu$m) emission, 
respectively, as a function of $T$. The symbols are the same as in the top panel.
}
   \label{fig:H2Oratio}
   \end{figure}

%
%__________________________________________________________________

\section{Analysis}
\label{sect:analysis}

In this section we discuss the excitation conditions of the observed lines,
complemented with Spitzer H$_2$ data, when available.
In particular, we will concentrate our analysis on the L1448-B2 shock, 
where we detected the largest number of lines with high signal-to-noise ratio ($S/N$). 
Given the observed strict correlation between H$_2$O and H$_2$ mid-IR emission,
we will first use the Spitzer H$_2$ lines to constrain the H$_2$O temperature.
The H$_2$O line ratios and absolute intensities will then be used to
derive the density and column density of the gas and the size of the emitting region.
The physical parameters derived for H$_2$O are then used for the CO emission
and both H$_2$O and CO abundances with respect to H$_2$ are estimated. 
Finally, the emission at the other shock spots with respect to the excitation 
conditions in L1448-B2 will be discussed.

\subsection{The L1448-B2 shock}
\label{subsect:B2spot}

\subsubsection{H$_2$ rotational diagram}
\label{subsubsect:H2diagram}

We used Spitzer H$_2$ observations by \citet{giannini2011} and converted the H$_2$ intensities, 
averaged in a 13$^{\prime\prime}$ beam, into column densities ($N_{\rm u}$) 
to construct the H$_2$ rotational diagram, which is presented in the upper panel of Fig.~\ref{fig:H2_rot-diagr}.
The fluxes have not been corrected for visual extinction, since it is only a marginal effect
\citep[$A_{\rm V} = 6$~mag for L1448][]{giannini2011,nisini2000}.
As explained in \citet{giannini2011}, the ortho-to-para ratio 
is temperature dependent: in particular, an ortho-to-para ratio close to 1 is found for the 
low-$J$ transitions that trace gas at $T \lesssim 400$~K, while the high temperature 
equilibrium value of 3 is reached by the lines with $J$ %$J_{up}$ 
larger than 3.

The fact that the observed transitions do not align on a single straight line on the %Boltzmann 
rotational diagram indicates that gas components at different temperatures are present within 
the spatial resolution element or along the line of sight.
In particular, two temperature components can be identified in the diagram, 
in the assumption of LTE conditions: 
a warm component at $T$ in the range $\sim 350-450$~K, where the uncertainty 
depends on the lines considered for the fit, i.e. the S(0)--S(2) or the S(0)--S(3) lines, 
and a hot component at $T \sim 1100$~K, by fitting the S(3) (or S(2)) to S(7) lines.
The H$_2$ column density is $N$(H$_2) \sim 0.8--1.2 \times 10^{20}$~cm$^{-2}$ 
for the warm component and $N$(H$_2) \sim 1.1 \times 10^{19}$~cm$^{-2}$ for the hot component.
In conclusion, the properties of the hot component are well defined from the H$_2$ rotational diagram, 
while a slightly larger range of parameters can be associated with the warm component.

\begin{table*}
\caption{Summary of the best-fit models derived for the two gas components at the L1448-B2 position.}
\label{table:models}
\centering
\begin{tabular}{l c c c c c c c }
\hline\hline
Comp. & $T$ & $n$(H$_2$) & $N$(H$_2$O) & $\Theta$  & $N$(CO) & [H$_2$O]/[H$_2$]\tablefootmark{a} & [CO]/[H$_2$]\tablefootmark{a} \\
& (K) & (cm$^{-3}$) & (cm$^{-2}$) & (arcsec)  &  (cm$^{-2}$) & & \\
\hline
Warm\tablefootmark{b} &  450 &                 10$^6$ &       $3 \times 10^{14}$ &             17 &       $3 \times 10^{15}$ &      $3-4  \times 10^{-6}$ &   $3-4 \times 10^{-5}$ \\
Hot                   & 1100 &  $(0.5-5) \times 10^6$ & $(0.4-2) \times 10^{16}$ & $\sim 1$ & $(1.5-3) \times 10^{16}$ & $(0.3-1.3) \times 10^{-5}$ & $(1-2) \times 10^{-5}$ \\      % $\Theta < 5$
\hline
\end{tabular}
\tablefoot{ 
\tablefoottext{a}{The H$_2$O and CO abundances of each gas component are obtained from the H$_2$O and CO column 
densities after correcting for the relative beam filling factor.}
\tablefoottext{b}{See the B2-2 model from \citet{santangelo2012}, shown in their Table~2. The H$_2$O column density 
is, however, slightly different because it has been derived using the collisional rate 
coefficients from \citet{dubernet2006,dubernet2009} and \citet{daniel2010,daniel2011}.}
}
\end{table*}

   \begin{figure*}%[ht]
   \centering
   \includegraphics[width=0.65\textwidth]{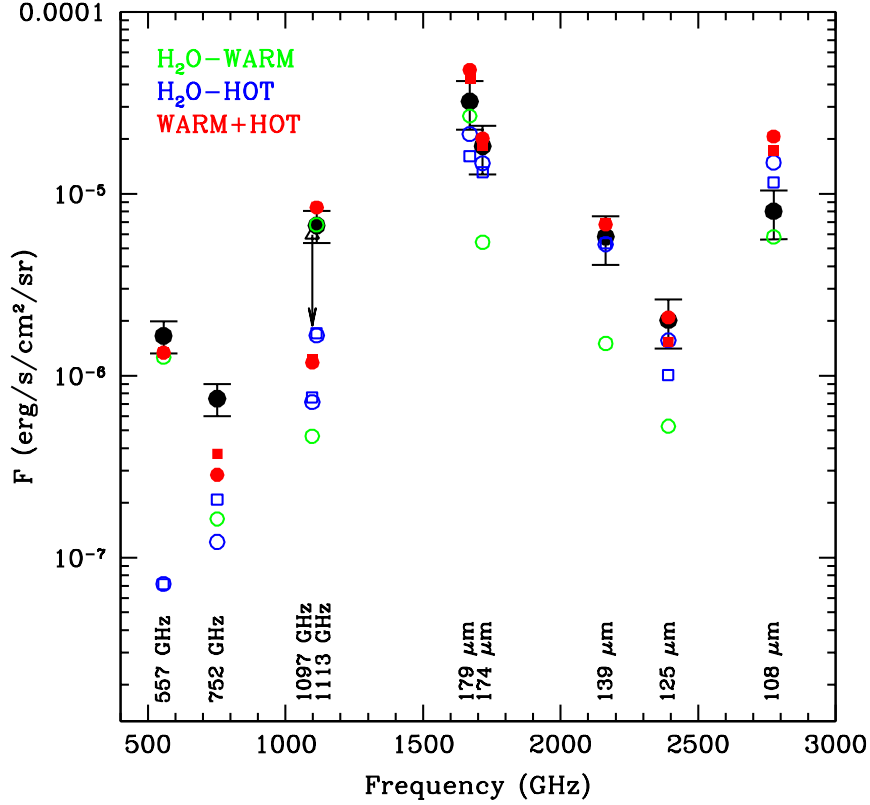}
   \caption{Comparison between the observed H$_2$O fluxes (black dots) and the two best-fit models 
for L1448-B2, which are given in Table~\ref{table:models}: the green model is the fit to 
the HIFI H$_2$O lines 
and the two blue models are the extremes of the obtained density range that fits the PACS H$_2$O lines (squares 
represent $n$(H$_2$)$=5 \times 10^5$~cm$^{-3}$ and circles represent $n$(H$_2$)$=5 \times 10^6$~cm$^{-3}$).
The red model represents the sum of the fluxes predicted for each line by the green and the two blue models.
The fluxes predicted by the models have been corrected for the relative predicted filling factors.
Calibration uncertainties of 20\% for the HIFI data and 30\% for the PACS data have been assumed.
The open triangle represents the upper limits of the HIFI H$_2$O $3_{12}-3_{03}$ line ($E_{\rm u} = 249$~K).
}
   \label{fig:HIFI+PACS}
   \end{figure*}

\subsubsection{The H$_2$O emission} 
\label{subsubsect:h2o}

Previous HIFI observations in L1448-B2 of H$_2$O lines with excitation energies $E_{\rm u}$ ranging 
from 53 to 137~K\footnote{The ortho-H$_2$O $3_{12}-3_{03}$ line with $E_{\rm u} = 249$~K
was not detected in B2 \citep[see][]{santangelo2012}, therefore the H$_2$O line with the highest energy used 
for the fit was the para-H$_2$O $2_{11}-2_{02}$ line with $E_{\rm u} = 137$~K.} 
\citep{santangelo2012} are consistent with very dense gas with $n$(H$_2$)$\sim 10^6$~cm$^{-3}$ 
and $T=450$~K and with moderate H$_2$O column densities of $\sim 3$~10$^{14}$~cm$^{-2}$. 
The bulk of the HIFI H$_2$O emission can be thus associated with the warm component identified from the H$_2$ rotational 
diagram.

To analyse the excitation conditions of the H$_2$O emission observed with PACS,
we used the radiative transfer code RADEX \citep{vandertak2007} in the plane-parallel geometry,
with the collisional rate coefficients from \citet{dubernet2006,dubernet2009} and \citet{daniel2010,daniel2011},
to build a grid of models with density ranging between 
10$^5$ and 10$^8$~cm$^{-3}$ and H$_2$O column density between 10$^{15}$ and 10$^{18}$~cm$^{-2}$.
We adopted a typical line width $\Delta v$ of 50~km~s$^{-1}$ (full-width at zero intensity, FWZI), 
from the spectrally resolved HIFI observations of H$_2$O \citep{santangelo2012}.
An uncertainty on the assumed line width value translates into an uncertainty on the H$_2$O column 
density determination, since the H$_2$O line ratios depend on the ratio $N$(H$_2$O)/$\Delta v$.
An ortho-to-para ratio equal to 3 was assumed, as implied by the HIFI observations of the warm component.

   \begin{figure*}%[ht]
   \centering
   \includegraphics[width=0.65\textwidth]{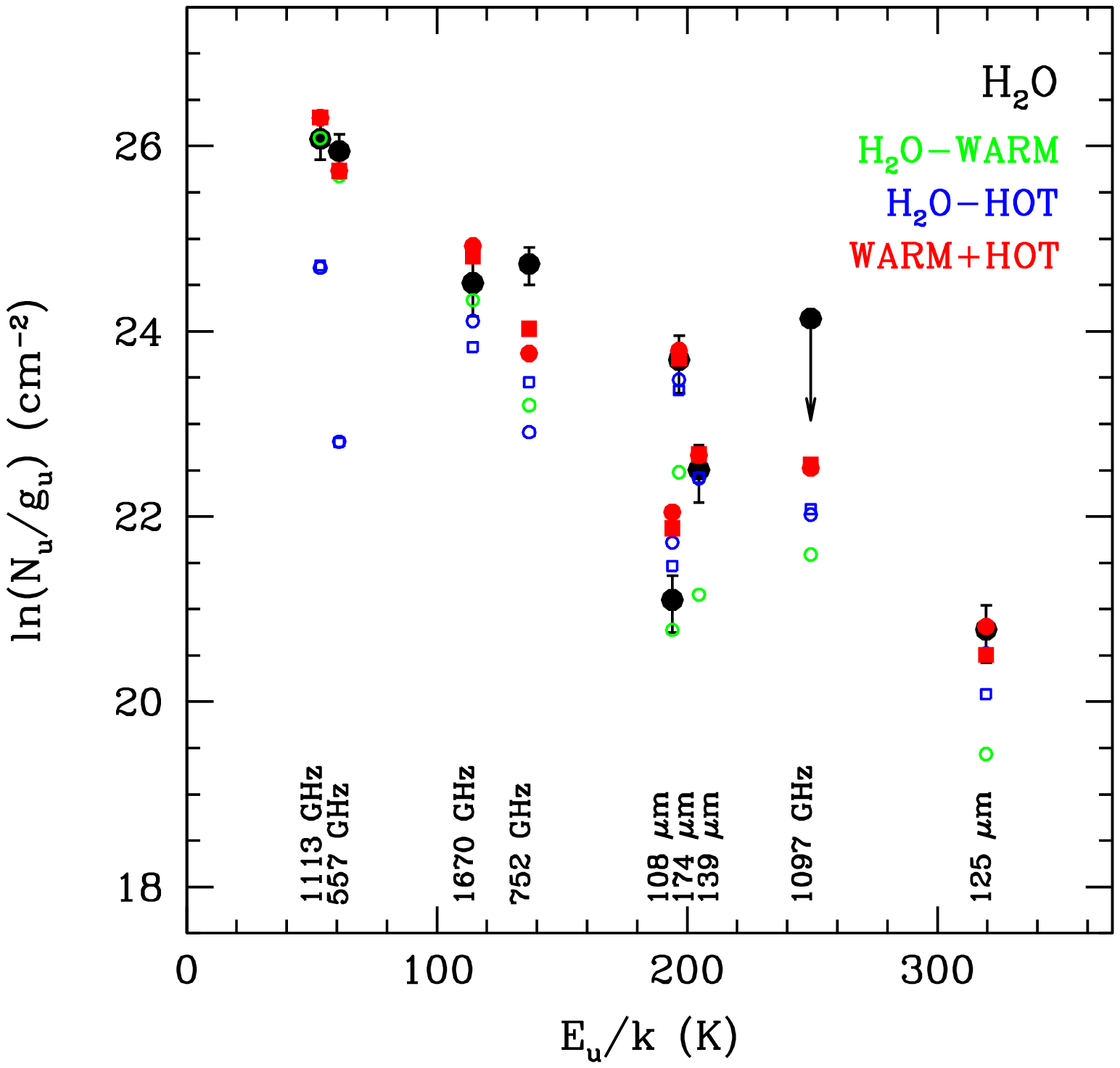}
   \caption{H$_2$O rotational diagram at L1448-B2.
Calibration uncertainties of 20\% for the HIFI data and 30\% for the PACS data have been assumed.
The predictions of the two best-fit models for L1448-B2, corrected for the relative predicted filling factors, 
are shown (see Table~\ref{table:models} and Fig.~\ref{fig:HIFI+PACS}). 
Symbols are as in Fig.~\ref{fig:HIFI+PACS}.
}
   \label{fig:H2O_rot-diagr}
   \end{figure*}

Figure~\ref{fig:H2Oratio} shows the ratios between H$_2$O lines observed with PACS 
(having higher excitation than those observed with HIFI) as a function of temperature, for two 
values of H$_2$ density and H$_2$O column density.
The warm component at $T \sim 450$~K does not reproduce the PACS H$_2$O line ratios.
In particular, the top panel shows the 179~$\mu$m/174~$\mu$m line ratio.
The arrows in the plot indicate that the observed ratio (shaded band) is an upper limit, 
since the H$_2$O $2_{12}-1_{01}$ line 
($E_{\rm u} \sim 114$~K) is more contaminated than the H$_2$O $3_{03}-2_{12}$ line ($E_{\rm u} \sim 197$~K)
by the warm gas component traced by the lower excitation H$_2$O lines observed with HIFI.
The low H$_2$O column density $N$(H$_2$O)$\sim 3 \times 10^{14}$~cm$^{-2}$, 
derived for the warm component from the HIFI H$_2$O observations, 
does not reproduce the 179~$\mu$m/174~$\mu$m line ratio.
This indicates the presence 
of an additional gas component with respect to the warm one traced by the bulk of the HIFI observations.
In particular, a hot component with temperature higher than 600~K
and H$_2$O column density larger than a few 10$^{15}$~cm$^{-2}$ 
is required to reproduce the higher excitation H$_2$O emission observed with PACS.

This evidence suggests that the bulk of the H$_2$O emission observed with PACS is associated with
the hot component which is seen in the H$_2$ rotational diagram.
Assuming a temperature $T = 1100$~K for this component from the H$_2$ rotational diagram 
(see Fig.~\ref{fig:H2_rot-diagr}) and H$_2$ density $n$(H$_2$)$\geq 10^5$~cm$^{-3}$, 
as obtained by \citet{giannini2011}, the density and column density of this component
are derived by fitting the intensity of all the PACS lines with excitation energy level 
$E_{\rm u} \gtrsim 190$~K and varying $n$(H$_2$), $N$(H$_2$O) and the size of the emitting region $\theta$.
Table~\ref{table:models} summarizes the results of the fit: 
unlike the warm component, the emission region of the hot component should be compact (a few arcsec). 
In particular, assuming an emitting size of 1 arcsec, 
the hot component requires a density $n$(H$_2$)$\sim 5 \times 10^5-5 \times 10^6$~cm$^{-3}$
and column density $N$(H$_2$O)$\sim 4 \times 10^{15}-2 \times 10^{16}$~cm$^{-2}$. 
The obtained column densities correspond to moderately optically thick lines (the optical depths of the 
H$_2$O lines are lower than $\sim 23$, corresponding to the maxim optical depth of the 179~$\mu$m line).

A visualization of the obtained results is presented in Fig.~\ref{fig:HIFI+PACS}, which 
shows the two separate models for the H$_2$O emission, i.e. the warm and hot components, 
and the sum of the two models in red.
The fluxes predicted by the models have been corrected for the 
filling factors (using $\theta_{\rm source}^2/(\theta_{\rm source}^2+\theta_{\rm beam}^2)$) 
obtained by assuming the emitting size derived from the excitation analysis.
Except for the H$_2$O $2_{21}-1_{10}$ line at 108.1~$\mu$m, which is 
over-estimated by a factor of 2.5, and the $2_{11}-2_{02}$ line at 752~GHz, 
which is under-estimated by a factor of 2, 
all H$_2$O lines are well reproduced by the two-component model.

We note that this model predicts that the hot component seen in L1448-B2 
should contribute very little to the emission
of the low-excitation H$_2$O lines observed with HIFI \citep{santangelo2012}
and indeed these lines show very similar profiles with no clear evidence of
variations in shape with excitation.
However, the HIFI observations of L1448-R4 and L1157-R 
show a different trend, with high velocity gas preferentially associated
with the low-excitation lines \citep[see][]{santangelo2012,vasta2012}.
Nevertheless, in these cases geometrical effects related to the presence
of bow shocks and self-absorption by cold H$_2$O gas in the lines at lower excitation 
may contribute to modifying the line profiles.

   \begin{figure*}%[ht]
   \centering
   \includegraphics[width=0.65\textwidth]{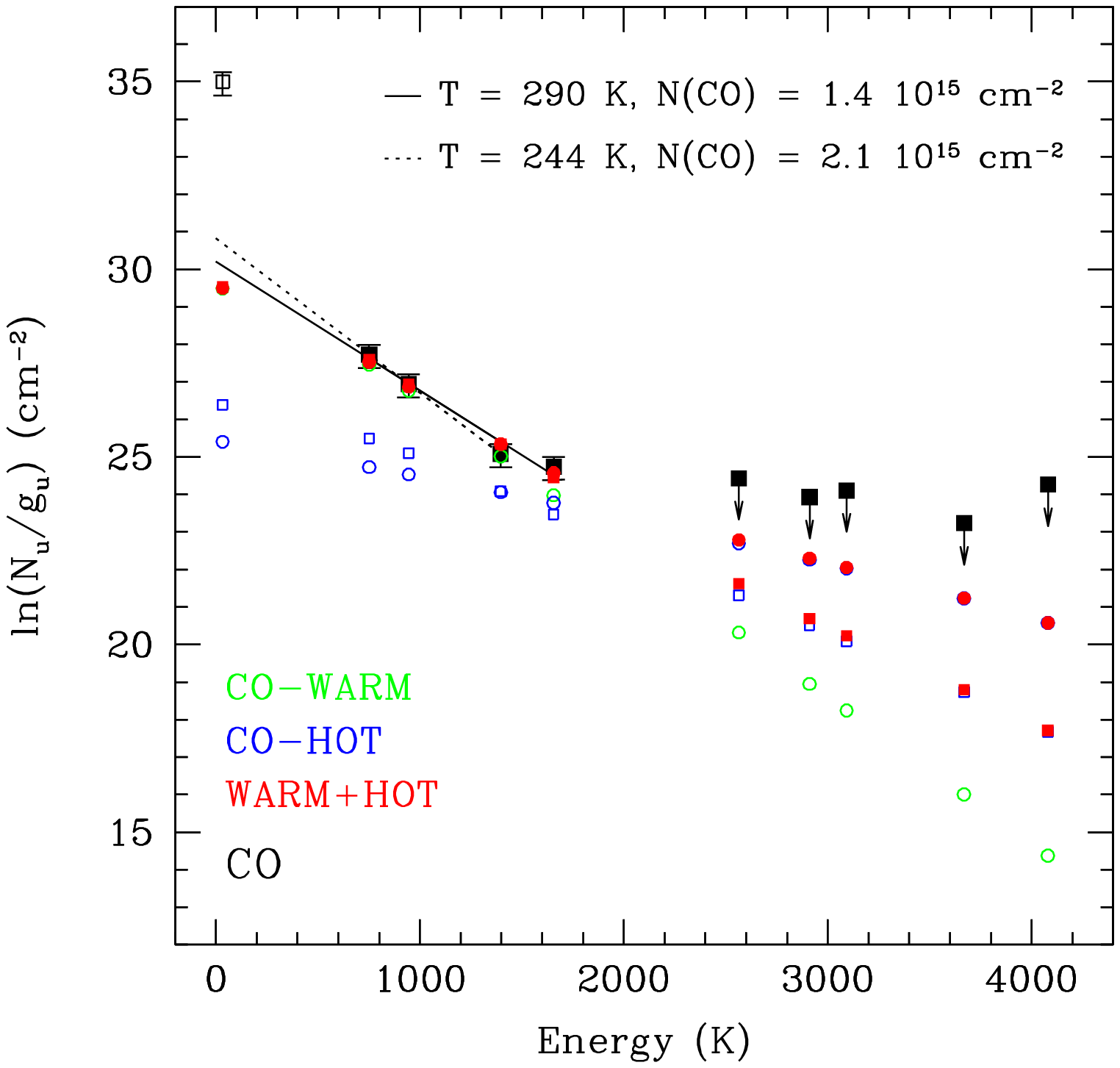} 
   \caption{Rotational diagram at L1448-B2 for the CO emission lines 
(both the detections and the non-detections) observed with PACS
(in a 12$.\!\!^{\prime\prime}$6 beam) and the JCMT CO($3-2$) line 
(empty symbol; beam size equal to 14$^{\prime\prime}$).
Calibration uncertainties of 30\% have been assumed.
The solid line represents the linear fit to the four detected CO lines, 
the dotted line the linear fit only to the three lower excitation CO lines.
As in Fig.~\ref{fig:H2O_rot-diagr}, the predictions of the two best-fit models for L1448-B2
corrected for the relative predicted filling factors are shown.
}
   \label{fig:CO_rot-diagr}
   \end{figure*}

Finally, Fig.~\ref{fig:H2O_rot-diagr} presents the H$_2$O rotational diagram, with the 
flux predictions from the two separate models for H$_2$O and from their sum.
The models identify two gas components in the rotational diagram, with the blue one
(associated with the hot component) showing more scatter than the green one 
(associated with the warm component), because of the larger associated optical depths.
However, when the sum of the two separate models is considered, the two temperature components 
are no longer discernible. The total rotational ladder shows a single-temperature aspect,
although the large scatter suggests that subthermal excitation and optical depth effects 
are significant.
The rotational temperature obtained from a single-temperature fit is $\sim 50$~K, 
which is within the range of rotational temperatures obtained for 
low-mass Class~0 protostars \citep[e.g.][]{herczeg2012,goicoechea2012,karska2013}.

\subsubsection{The CO emission}
\label{subsubsect:co}

Figure~\ref{fig:CO_rot-diagr} shows the CO rotational diagram obtained 
by converting the fluxes of the high-$J$ CO lines observed with PACS 
and the CO($3-2$) line observed with the JCMT telescope (beam size equal to 14$^{\prime\prime}$) 
into column densities ($N_{\rm u}$). 
A global fit to the PACS CO lines reveals 
a gas with rotational temperature 
$T \sim 290$~K and CO column density, 
averaged in the 12$.\!\!^{\prime\prime}$6 beam, of $N$(CO)$\sim 10^{15}$~cm$^{-2}$. 
This is consistent 
with the warm gas component identified from the H$_2$ rotational diagram and associated with the bulk 
of the H$_2$O emission observed with HIFI.
Although only four CO lines have been detected with PACS, 
a hint of a possible curvature occurs at excitation energies $E_{\rm u} \geq 1000$~K, 
since the CO($24-23$) transition lies above the straight line followed by the other PACS lines. 
If we assume that this line comes from a different gas 
component, a slightly lower temperature $T \sim 240$~K is found from the lower excitation
PACS lines and correspondingly $N$(CO)$\sim 2 \times 10^{15}$~cm$^{-2}$.
The same diagram shows that the CO($3-2$) line lies well above the other CO lines, 
which is consistent with its origin in a colder gas.

   \begin{figure*}%[ht]
   \centering
   \includegraphics[width=0.65\textwidth]{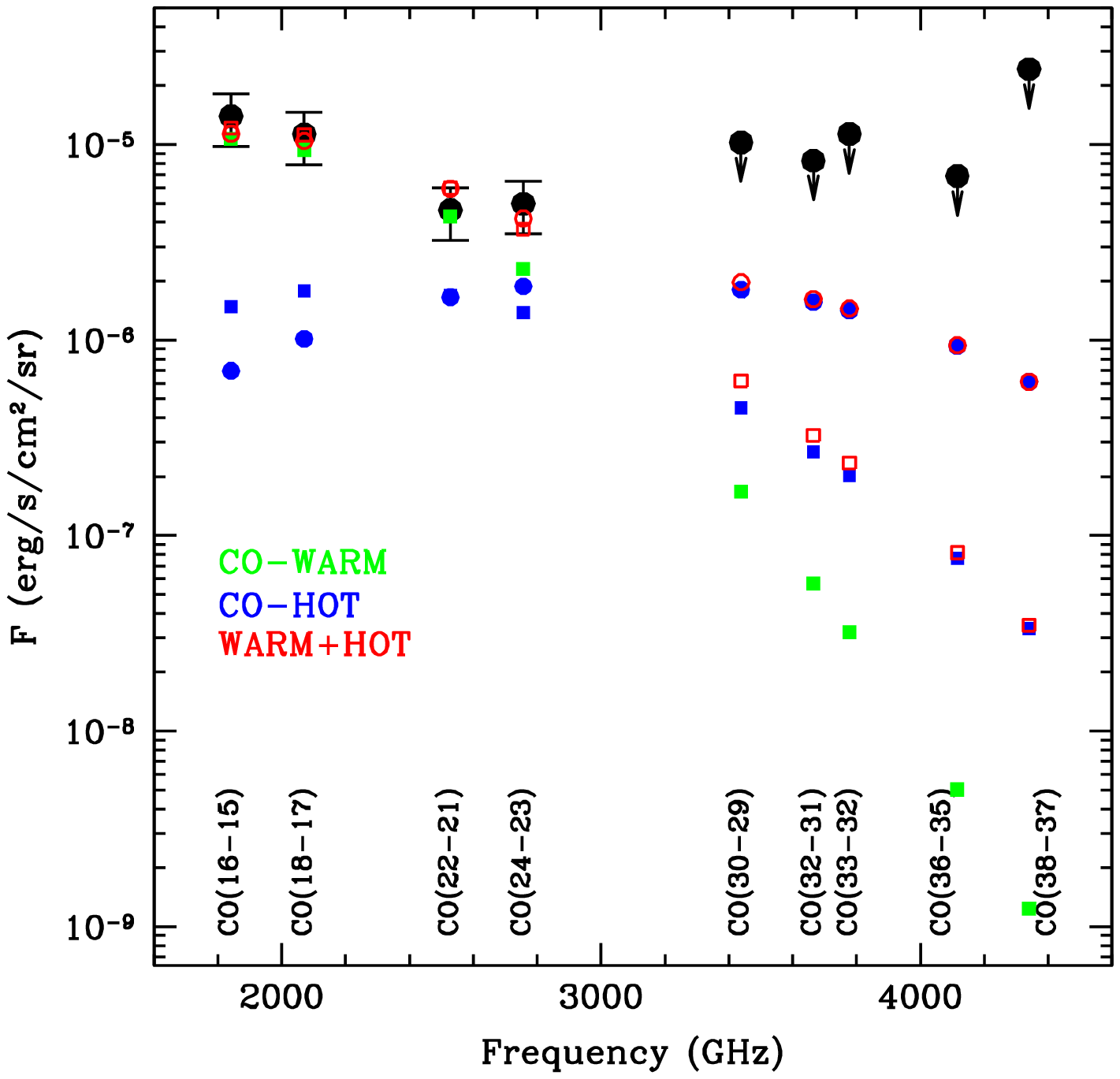}
   \caption{The same comparison presented in Fig.~\ref{fig:HIFI+PACS}, but for the CO fluxes 
measured with PACS toward L1448-B2 (both the detections and the non-detections). The two blue models represent 
the extremes of the density range derived from the H$_2$O excitation analysis.}
   \label{fig:HIFI+PACS_CO}
   \end{figure*}

The presence of multiple excitation temperature components in the CO emission
has been found by other studies of CO ladders 
in low-mass Class~0 protostars and their outflows
\citep[see e.g.][]{vankempen2010b,benedettini2012,goicoechea2012,herczeg2012,yildiz2012,yildiz2013,karska2013,manoj2013}.
In particular, \citet{karska2013} present CO rotational diagrams for a large sample of 
protostars, showing two distinct components, a warm component with $T_{\rm rot} \sim 300$~K 
and a hot component with $T_{\rm rot} \sim 700$~K, in addition to a cold component 
with $T_{\rm rot} \sim 100$~K, observed in the $J \lesssim 14$ lines \citep{goicoechea2012,yildiz2012,yildiz2013}.
They found the break between warm and hot gas in the CO diagrams around $E_{\rm u} \sim 1500$~K.
Thus, the presence of different components in our PACS CO data, a warm and a hot 
component, is probably valid and may reflect true differences in the excitation conditions of 
the gas traced by the different ranges of CO transitions in Class 0 sources.

We can then use the physical conditions derived for the warm and hot H$_2$O components 
to verify whether they are able to reproduce our 
PACS CO observations. The comparison is presented in Fig.~\ref{fig:HIFI+PACS_CO}.
In particular, we used the temperature and density 
derived from the H$_2$O analysis 
to fit the CO line ratios, normalizing the warm component to the CO($16-15$) line and 
deriving the CO column density of the hot component so that the sum of the 
two components (warm plus hot) reproduced the observed CO fluxes and upper limits.
The CO column densities derived in this fashion are reported in the last column of 
Table~\ref{table:models}.
The two blue models in Fig.~\ref{fig:HIFI+PACS_CO} for the hot gas component represent the extremes 
of the density range obtained from the H$_2$O excitation analysis 
(see Table~\ref{table:models}). We obtained $N$(CO$)=(1.5-3) \times 10^{16}$~cm$^{-2}$
for the hot gas component and $N$(CO$)=3 \times 10^{15}$~cm$^{-2}$ for the warm component, 
both averaged over the relative emitting size.

To summarize, the CO and H$_2$O line ratios trace two gas components, 
a warm gas component at $T \sim 450$~K (with $n$(H$_2$)$=10^6$~cm$^{-3}$), which is visible in the 
H$_2$O emission with $E_{\rm u}= 53-137$~K and the PACS CO data up to $J=22-21$, 
and a hot gas component at $T \sim 1100$~K (with $n$(H$_2$)$=(0.5-5) \times 10^6$~cm$^{-3}$), 
which is traced by the H$_2$O observations with $E_{\rm u} > 190$~K and the PACS higher-$J$ CO emission. 
These two gas components are associated with a warm and hot component, respectively, in the Spitzer mid-IR H$_2$  emission.

\subsubsection{H$_2$O and CO abundance ratios}
\label{subsubsect:abundance}

A direct estimate of the H$_2$O and CO abundances with respect to H$_2$
for both gas components can be obtained by comparing the 
column density of these species, averaged over a 13$^{\prime\prime}$ beam.
We find an [H$_2$O]/[H$_2$] abundance ratio of $(3-4) \times 10^{-6}$ for the warm component 
and $(0.3-1.3) \times 10^{-5}$ for the hot component.
The inferred H$_2$O abundances are much higher than the typical value of 
$\sim 10^{-9}-10^{-8}$, which is found in cold interstellar clouds \citep[e.g.][]{caselli2010}.
However, even for the hot component, this is lower than $\sim 10^{-4}$, 
which is the value expected in hot shocked gas \citep[e.g.][]{kaufman1996,flower2010}.
The derived H$_2$O abundances for the warm component are consistent with the values 
obtained by \citet{santangelo2012,vasta2012,nisini2013} from HIFI velocity-resolved observations.
In particular, \citet{nisini2013}, from the analysis of H$_2$O $2_{12}-1_{01}$ and $1_{10}-1_{01}$ maps 
of L1448, derived a relatively constant water abundance along the outflow of about $(0.5-1) \times 10^{-6}$, 
with an increase by roughly one order of magnitude at the protostar position.
Similarly, low H$_2$O abundances in the warm gas have been derived in other outflows
by several authors \citep[e.g.][]{bjerkeli2012,tafalla2013}.

On the other hand, we derive a [CO]/[H$_2$] abundance of $(3-4) \times 10^{-5}$
for the warm component and $(1-2) \times 10^{-5}$ for the hot component. 
The derived [CO]/[H$_2$] abundances do not depend on the emitting size, 
because both CO and H$_2$ lines are optically thin; therefore, their absolute intensity 
depends on the beam diluted column density.

Our data suggest that the CO abundance is lower by a factor from 3 to 10 
with respect to the canonical value of $2.7 \times 10^{-4}$ measured for dense interstellar
clouds \citep[e.g.][]{lacy1994}.
Shocks that are non-dissociative, like those implied by our molecular 
observations (see Sect.~\ref{sect:shock_models}), are not expected to alter the original CO/H$_2$ 
abundance ratio in the cloud. We point out however that a CO abundance less than the canonical
value has been recently measured in different environments, including 
the inner envelopes of low- and intermediate-mass protostars \citep[e.g.][]{yildiz2010,yildiz2012,fuente2012} 
and toward the Orion region \citep[][]{wilson2011}, which indicates that such low values 
are indeed not peculiar to the considered shock regions.

By comparing the H$_2$O and CO column densities, we find a [H$_2$O]/[CO] abundance ratio 
of 0.1 for the warm and $0.1-1.3$ for the hot component.
Thus, an estimate of the H$_2$O abundance, based on the assumption 
that the CO abundance with respect to H$_2$ is equal to 10$^{-4}$, would lead to higher values for the hot component 
(about $1-13$~10$^{-5}$).
For this reason our obtained H$_2$O abundance values are different
from those obtained previously from ISO observations 
\citep[e.g.][]{nisini1999,nisini2000,giannini2001}: 
our analysis points to a CO abundance with respect to H$_2$ lower than the standard value of 10$^{-4}$
for the hot component and correspondingly to a lower H$_2$O abundance.

\subsubsection{The spatial extent of the warm and hot components}

According to the excitation analysis, 
different sizes are associated with the two H$_2$O gas components: the warm gas is 
found to be rather extended (17$^{\prime\prime}$), while the hot gas should be compact ($< 5^{\prime\prime}$).
Based on our model, from Fig.~\ref{fig:HIFI+PACS} we expect
the contribution of the warm component to the total H$_2$O flux at 179~$\mu$m to be similar or stronger than 
that of the hot component, whereas at 174~$\mu$m the hot component dominates the H$_2$O emission
with little contribution from the warm component.
One way of studying the spatial extent of the two components and verifying the results obtained 
from our analysis is to use the maps of these two H$_2$O lines (179 and 174~$\mu$m), 
which are also the strongest H$_2$O lines we detected with PACS, 
and analyse the relative contribution of the two predicted components from their line ratio.
In particular, we expect the ratio between the 179~$\mu$m and the 174~$\mu$m H$_2$O lines 
to increase going from the central position outwards, thanks to the dominant contribution 
of the compact central component to the H$_2$O 174~$\mu$m flux.

\begin{figure}%[ht]
   \centering
   \includegraphics[angle=-90,width=0.48\textwidth]{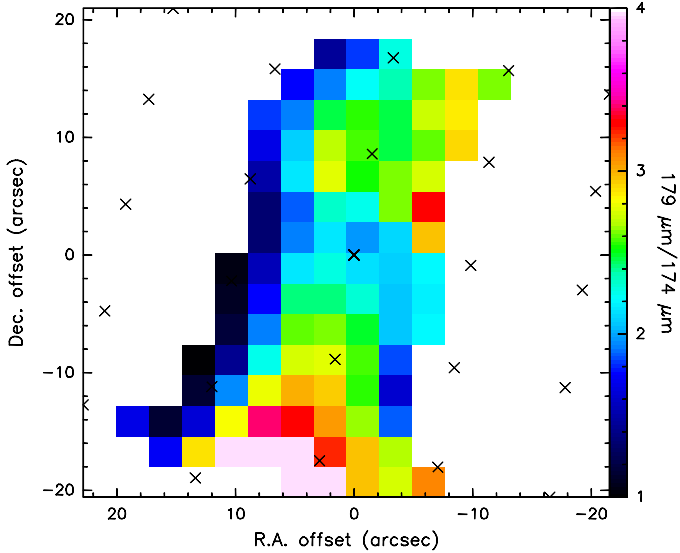}
   \caption{Ratio between the H$_2$O $2_{12}-1_{01}$ map at 179~$\mu$m and the 
$3_{03}-2_{12}$ map at 174~$\mu$m for L1448-B2. 
The ratio is shown only above a 5~$\sigma$ detection level in both maps.
The crosses represent the pointing of the 25 spaxels.
}
   \label{fig:179su175}
   \end{figure}

Figure~\ref{fig:179su175} presents the ratio between the PACS maps of the two H$_2$O lines (at 179~$\mu$m 
and 174~$\mu$m). 
As predicted by our excitation analysis, the H$_2$O line ratio increases going from the centre 
of the map toward the edges in both directions along the outflow. 
This result supports the scenario in which two gas components coexist: a compact component 
which dominates the H$_2$O emission above $E_{\rm u} \sim 190$~K 
and an extended component that dominates the H$_2$O emission at lower excitation energies.

\subsection{Excitation conditions and water abundance at the other shocked positions}
\label{subsect:other_spots}

At the other selected shock spots a detailed analysis like the one we performed 
for L1448-B2 is precluded because of the smaller number of lines, the lower $S/N$ of the detections, 
and the lack of H$_2$ data that would allow us to get a direct measure of the water abundance.
The only other position, among the selected ones, where Spitzer spectroscopic
data are available is L1157-R \citep{nisini2010b}. 
Therefore, at this position an estimate of the water abundance can be obtained in a similar fashion.

The rotational diagram, constructed in L1157-R from the Spitzer mid-IR data 
(see lower panel of Fig.~\ref{fig:H2_rot-diagr}), shows once more the presence of two 
gas components: a warm component at a temperature of about $T \sim 420$~K and a hot 
component with $T \sim 800-850$~K. The break is found at approximately 2000~K.
The corresponding H$_2$ column densities are $N$(H$_2) \sim 1.2 \times 10^{20}$~cm$^{-2}$ 
for the warm component and $N$(H$_2) \sim (0.7-1.3) \times 10^{19}$~cm$^{-2}$ for the hot 
component, both averaged over 13$^{\prime\prime}$.

As we did for L1448-B2, we assume that the bulk of the H$_2$O emission observed with 
PACS is associated with the hot component identified from the H$_2$ rotational diagram,
and we adopt a RADEX analysis to derive the excitation conditions of this hot component. 
In particular, we derived the H$_2$ density, the H$_2$O column density, and the emitting size 
by fitting the PACS H$_2$O lines with excitation energy level $E_{\rm u} \gtrsim 190$~K, 
under the assumption of a temperature $T \sim 800-850$~K from the H$_2$ data and 
a line width of 25~km~s$^{-1}$ from the HIFI observations by \cite{vasta2012}. 

A compact gas component is found to be associated with the bulk of the PACS emission, 
with $n$(H$_2$)$\sim (0.1-5) \times 10^6$~cm$^{-3}$ and $N$(H$_2$O)$\sim (0.2-6) \times 10^{16}$~cm$^{-2}$.
The obtained excitation conditions appear to be similar to those derived for the hot 
component at L1448-B2, but a larger uncertainty on the H$_2$O column densities 
is associated with L1157-R. The corresponding abundance, obtained by comparing the 
H$_2$O column densities (corrected for the relative filling factor) with the H$_2$ column density, 
is in the range $(0.1-5) \times 10^{-5}$ (see Sect.~\ref{subsubsect:abundance} for comparison).

\begin{figure}%[ht]
   \centering
   \includegraphics[width=0.48\textwidth]{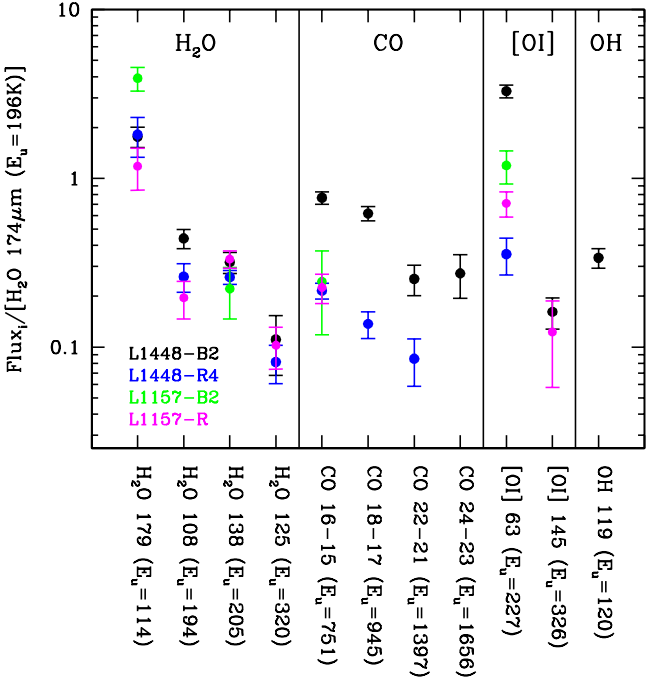}
   \caption{Line ratios between the H$_2$O, CO, [O{\sc i}], and OH lines and the H$_2$O $3_{03}-2_{12}$ line 
at 174~$\mu$m at all selected shock positions. 1~$\sigma$ errors are indicated with errorbars.
}
   \label{fig:line_ratios_otherSPOTS}
   \end{figure}

For the remaining two shock spots, namely L1448-R4 and L1157-B2, 
the lack of mid-IR H$_2$ data in both cases 
does not allow us to get constraints on the temperature and to estimate the H$_2$O abundance.
To investigate the physical and excitation conditions of the hot gas component at these positions,
we used the L1448-B2 shock position as a template and  
compared the ratios of the detected lines with the relative line ratios observed in L1448-B2.
The comparison is presented for all the selected shock spots in Fig.~\ref{fig:line_ratios_otherSPOTS}, 
where all the line ratios are normalized with respect to the H$_2$O $3_{03}-2_{12}$ line at 174~$\mu$m.
The observed H$_2$O line ratios are roughly comparable within the relative errors 
with those observed in L1448-B2, within a factor of 2.
We can thus conclude that the excitation conditions of the hot gas component 
are comparable in all selected shock positions, as already deduced for L1157-R.
We note that the bright H$_2$O 179/174$\mu$m line ratio at the L1157-B2 shock position may provide evidence 
for an older shock with respect to the other selected positions. Because this line ratio is indicative of the relative 
contribution between the warm and the hot component, the high value observed at L1157-B2  
may suggest a smaller contribution of the hot component relative to the warm component compared to the other shock 
positions. 
This is consistent with this position being the signpost of an older shock, as already suggested by previous studies 
\citep[e.g.][]{bachiller1997,rodriguezfernandez2010,vasta2012}.
Assuming a constant shock propagation velocity, \citet{gueth1998} derived a dynamical age for the L1157-B2 shock 
spot of $\sim 3000$~yr, which is larger than or similar to the typical cooling time of J-type and C-type shocks 
\citep[$\lesssim 10^2-10^3$~yr, respectively,][]{flower2010}. 
Therefore, in L1157-B2 the hot component has already had the time to cool down to a few hundred Kelvin.

   \begin{figure}%[ht]
   \centering
   \includegraphics[width=0.48\textwidth]{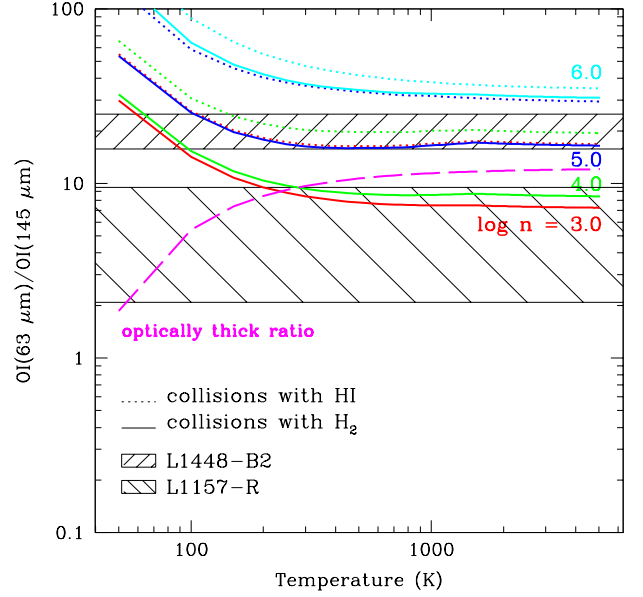}
   \caption{Optically thin [O{\sc i}]63~$\mu$m/[O{\sc i}]145~$\mu$m flux ratios 
as a function of temperature are shown 
in dotted lines for collisions with atomic hydrogen H 
and in solid lines for collisions with molecular hydrogen H$_2$ \citep[as in][]{liseau2006}.
The logarithms of the density (in cm$^{-3}$) are indicated for each curve. 
The broken line outlines the ratio of optically thick lines. 
Observed line ratios are depicted by the shaded areas for the L1448-B2 position and the L1157-R position.
The data have been smoothed to a common angular resolution of 12$.\!\!^{\prime\prime}$6.
}
   \label{fig:oIratio}
   \end{figure}

Figure ~\ref{fig:line_ratios_otherSPOTS} shows that CO/H$_2$O line ratios 
lower by a factor of $\sim 4$ with respect to L1448-B2 
are found at all other positions.
Under the assumption of similar H$_2$O excitation conditions, 
this would suggest a higher H$_2$O abundance with respect to L1448-B2,
which is in line with the range estimated for L1157-R using Spitzer mid-IR H$_2$ data. 

The L1448-R4 and L1157-B2 shock positions thus appear to be more similar to 
L1157-R than to L1448-B2 in terms of H$_2$O abundance.
This conclusion is supported by the PACS detection at L1448-B2 of OH and brighter [O{\sc i}] emission 
(see Fig.~\ref{fig:line_ratios_otherSPOTS}), 
which suggests either that not all oxygen has been converted into H$_2$O or that 
water is partially dissociated.
Indeed, the L1448-B2 shock position is intrinsically peculiar with respect 
to all the other selected positions, since it is close to the driving outflow source. 
Thus, this position may be affected by the strong UV radiation field coming from the central protostar
or from dissociative internal jet shocks \citep[e.g.][]{hollenbach1989,vankempen2009}, which can 
photodissociate the freshly formed H$_2$O.

\subsection{The [O{\sc i}] ratio}
\label{subsect:oIratio}

It is useful to compare the observed ratio between the [O{\sc i}] $^3P_1 - ^3P_2$ line at 63.2~$\mu$m 
and the [O{\sc i}] $^3P_0 - ^3P_1$ line at 145.5~$\mu$m to infer additional information 
on the gas excitation conditions \citep[see][]{liseau2006}. 
We detected both [O{\sc i}] lines 
only at the L1448-B2 and L1157-R positions (at the latter position the [O{\sc i}] line at 145~$\mu$m
was detected only at $\sim 2$~$\sigma$ level) 
and the measured [O{\sc i}]63/145$\mu$m ratios are $\sim 20$ 
and $\sim 6$, respectively.
The observed line ratios are displayed in Fig.~\ref{fig:oIratio}, along with the  
line ratios predicted from the RADEX code assuming optically thin lines, 
for collisions with atomic hydrogen or with molecular hydrogen; 
the predicted line ratio for optically thick lines is also shown.
We have neglected O excitation due to collisions with electrons, 
since it becomes relevant (i.e. it contributes more than 10\%) 
only for $n(e)/n({\rm H})$ fractions larger than 0.6, clearly in contrast 
with the mostly molecular/atomic gas observed in the considered shocks. 
Thus, assuming optically thin lines excited by collisions with H$_2$, the 
ratio observed at L1448-B2 is consistent with  
a H$_2$ volume density between $10^5$ and a few $10^5$~cm$^{-3}$ and a temperature $T \gtrsim 100$~K, 
which is within the range of parameters derived from
our excitation analysis (see Sects.~\ref{subsubsect:h2o} and \ref{subsubsect:co}).
On the other hand, assuming collisions with H, it corresponds to 
$n$(H)$\sim 10^3-10^4$~cm$^{-3}$ and $T \gtrsim 100$~K.
We can thus distinguish two possibilities for the origin of the [O{\sc i}] emission at L1448-B2:
either H$_2$O, CO, and [O{\sc i}] emission arise from the same molecular gas with 
density $n$(H$_2$)$\sim 5 \times 10^5$~cm$^{-3}$, or  
the [O{\sc i}] emission originates in a low-density component of atomic gas.
This will be discussed further in Sect.~\ref{sect:shock_models}.
Instead, the lower line ratio measured at L1157-R
is consistent either with $n$(H$_2$)$\sim 10^3-10^4$~cm$^{-3}$ and $T \gtrsim 200$~K for 
optically thin lines excited by collisions with H$_2$, 
or with optically thick lines and temperatures lower than 200~K.

Finally, in both positions the [O{\sc i}] column density averaged over the PACS beam
is of the order of 
$2-5$~10$^{15}$~cm$^{-2}$ at L1448-B2 and $5-10$~10$^{15}$~cm$^{-2}$ at L1157-R, 
which is consistent with optically thin lines.

   \begin{figure}%[ht]
   \centering
   \includegraphics[width=0.46\textwidth]{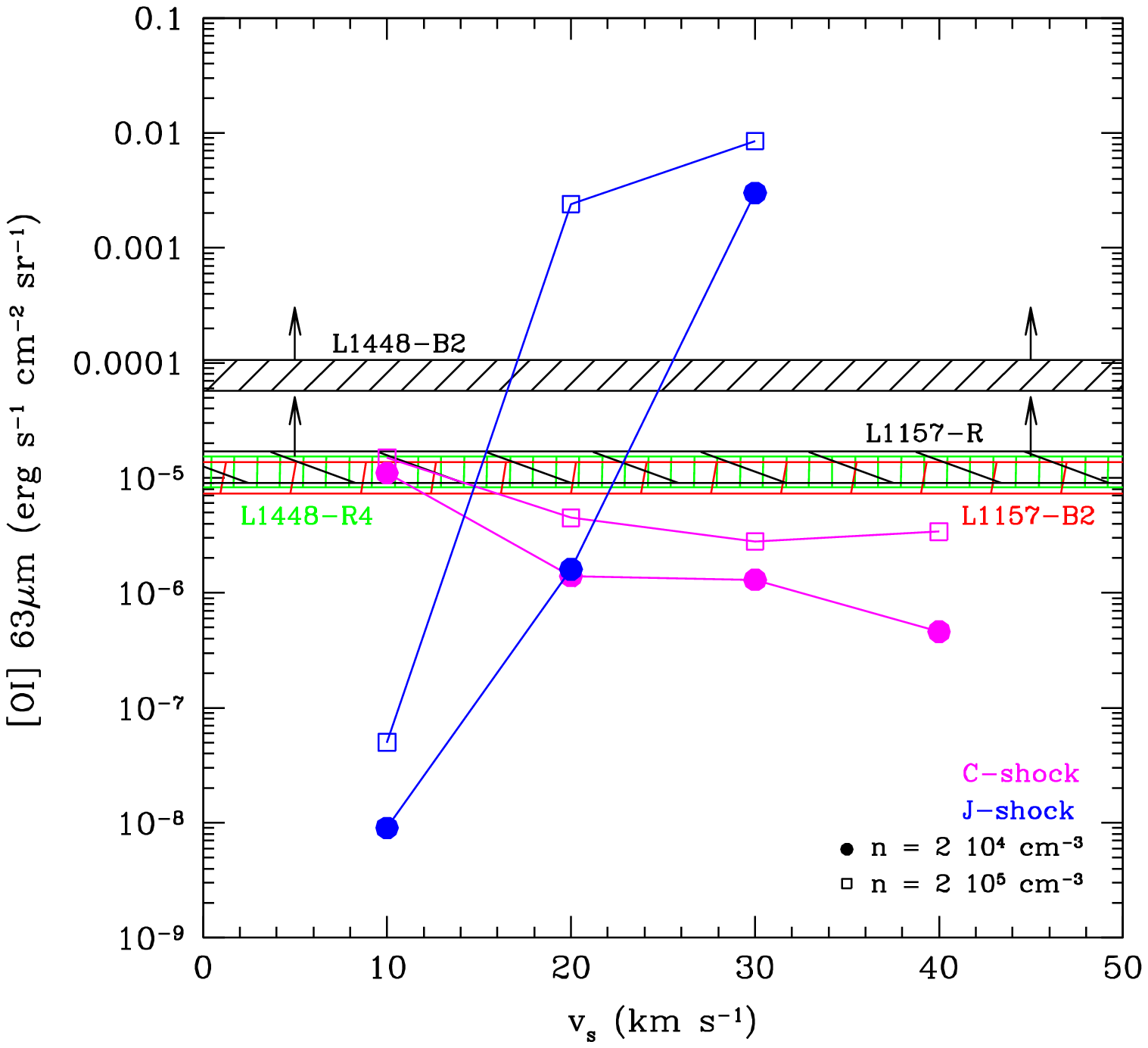}
   \includegraphics[width=0.46\textwidth]{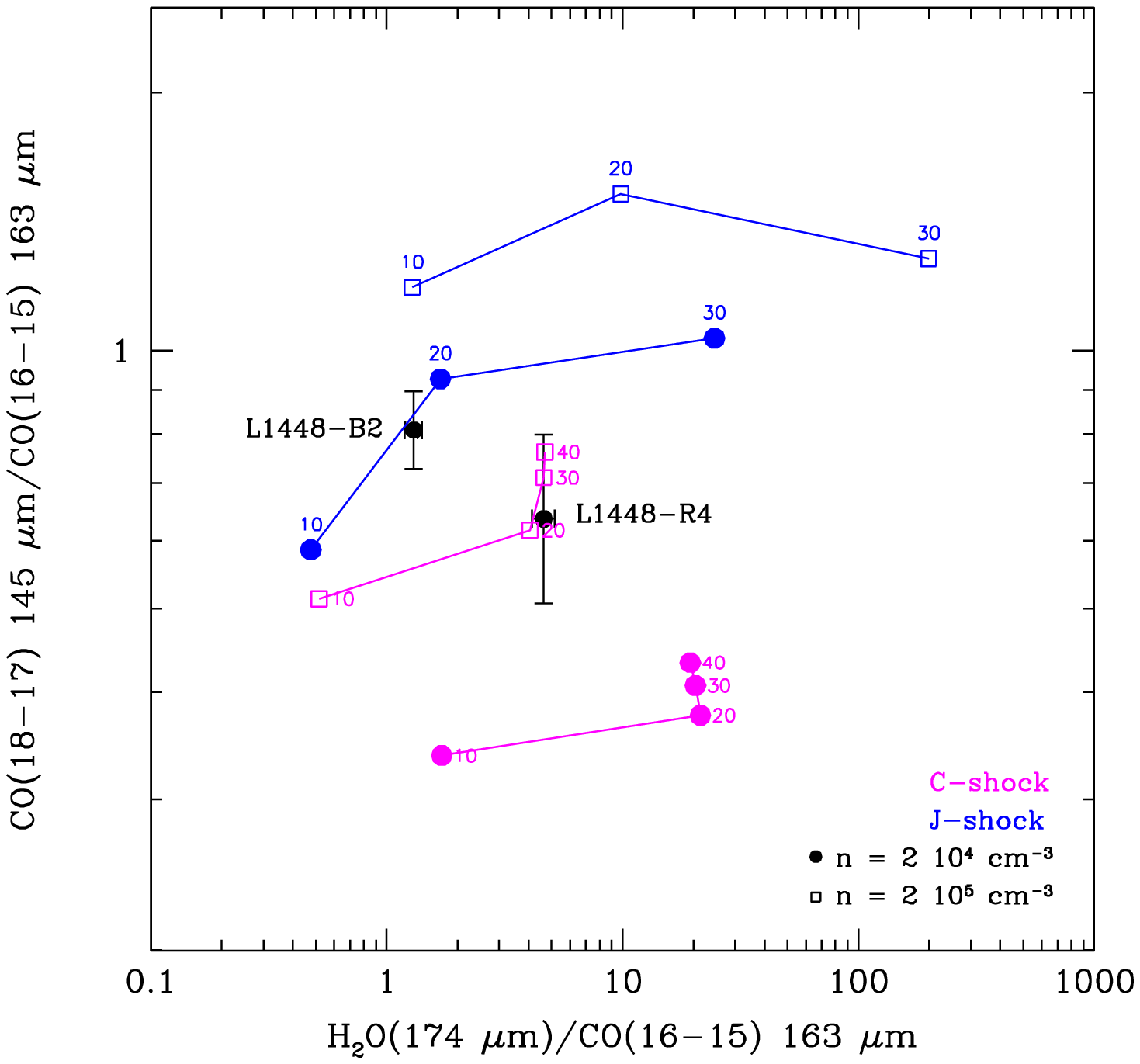}
   \caption{\emph{Upper panel}: Comparison between the [O{\sc i}] $^3P_1 - ^3P_2$ (63.2~$\mu$m) flux observed 
at the investigated shock positions (shaded bands) and the corresponding theoretical values predicted by 
the \citet{flower2010} shock models for C-type shocks (magenta) and J-type shocks (blue), 
as a function of the shock velocity in units of km~s$^{-1}$.
The fluxes measured in the central spaxel of the PACS maps have been used without smoothing to a common angular resolution.
Calibration uncertainties of 30\% have been assumed.
The arrows in the plot indicate that the absolute fluxes have to be considered as 
lower limits; they are beam diluted because we do not resolve the emitting size of the shock.
\emph{Lower panel}: Same comparison as in the upper panel, but 
for CO and H$_2$O line ratios. The observed values are depicted as black dots and the errorbars represent 1~$\sigma$ errors.
The data have been smoothed to a common angular resolution of 12$.\!\!^{\prime\prime}$6.
}
   \label{fig:shock-models_oI}
   \end{figure}

\section{Comparison with shock models}
\label{sect:shock_models}

The observed fluxes are compared
with the grid provided by \citet{flower2010} for stationary C- and J-type shock models.
The grid explores a range of shock velocities from 10 to 40~km~s$^{-1}$ and two pre-shock 
densities, $2 \times 10^4$ and $2 \times 10^5$~cm$^{-3}$.
In the upper panel of Fig.~\ref{fig:shock-models_oI} we present the observed flux of the [O{\sc i}] $^3P_1 - ^3P_2$ line 
at 63.2~$\mu$m with the shock model predictions.
Unsmoothed peak line fluxes have been used to minimize beam dilution effects.
At the L1448-B2 position only a 
J-type shock, with velocity $v_{\rm s} > 20$~km~s$^{-1}$ 
for pre-shock density $n = 2 \times 10^4$~cm$^{-3}$ and $v_{\rm s} > 10$~km~s$^{-1}$ for $n = 2 \times 10^5$~cm$^{-3}$, 
can reproduce the observed flux; C-type shocks under-estimate this line by at least one order of magnitude. 
A pre-shock density $n = 2 \times 10^5$~cm$^{-3}$, and a corresponding shock velocity 
$v_{\rm s} > 10$~km~s$^{-1}$, are not consistent with the results of the H$_2$O and CO excitation analysis 
(Table~\ref{table:models}).
From the comparison between this pre-shock density and the maximum post-shock density 
that can be evinced from the [O{\sc i}] line ratio (see Sect.~\ref{subsect:oIratio} and Fig.~\ref{fig:oIratio}), 
a very small compression factor would be derived. 
In addition, a comparison with shock models by \citet{hollenbach1989} shows that even a lower pre-shock density 
of 10$^{3}$~cm$^{-3}$ and shock velocity $v_{\rm s} \gtrsim 30$~km~s$^{-1}$ can reproduce our [O{\sc i}] data.
This suggests that at L1448-B2 the [O{\sc i}] emission originates  
in a fast dissociative shock with pre-shock density $n \lesssim 2 \times 10^4$~cm$^{-3}$ and $v_{\rm s} > 20$~km~s$^{-1}$.
The presence of a dissociative shock giving rise to ionizing photons is also supported 
by the detection at the [O{\sc i}] peak of OH at 119~$\mu$m and [Fe{\sc ii}] at 26~$\mu$m  
(see Fig.~\ref{fig:l1448mappe}).
On the other hand, at the other shock positions we are not able to discriminate between 
C- and J-type shocks. The observations are consistent either with a low-velocity 
($v_{\rm s} < 20$~km~s$^{-1}$) C-type shock or with a J-type shock with velocity 
$v_{\rm s} > 20$~km~s$^{-1}$ for $n = 2 \times 10^4$~cm$^{-3}$ and $v_{\rm s} > 10$~km~s$^{-1}$ 
for $n = 2 \times 10^5$~cm$^{-3}$.

\begin{table*}
\caption{Origin of the emission observed with PACS.}
\label{table:shock}
\centering
\begin{tabular}{l | c | c }
\hline\hline
Position  & [O{\sc i}] \& OH & H$_2$O \& high-$J$ CO \\
\hline
L1448-B2 & J-type shock ($v_{\rm s} > 20$~km~s$^{-1}$, $n\lesssim 2\times 10^4$~cm$^{-3}$) & J-type shock ($v_{\rm s} \lesssim 20$~km~s$^{-1}$, $n=2\times 10^4$~cm$^{-3}$) \\
L1448-R4 & J-type shock ($v_{\rm s} > 10$~km~s$^{-1}$)                                     & C-type shock ($v_{\rm s} > 20$~km~s$^{-1}$, $n=2\times 10^5$~cm$^{-3}$) \\
\hline
\end{tabular}
\end{table*}

A comparison of the observed CO and H$_2$O emission with shock models is presented in the lower panel of 
Fig.~\ref{fig:shock-models_oI} for the L1448-B2 and L1448-R4 shock positions.
At the L1448-B2 position the CO and H$_2$O emissions are also consistent with a J-type shock having a pre-shock 
density $n = 2 \times 10^4$~cm$^{-3}$, but 
at a lower shock velocity ($\lesssim 20$~km~s$^{-1}$) with respect to the [O{\sc i}] emission.
According to \citet{flower2010}, a high compression factor of about 100 is predicted 
for a low-velocity ($v_{\rm s} \lesssim 20$~km~s$^{-1}$) J-type shock, as observed at this shock position. 
This corresponds to a post-shock density of about $2 \times 10^6$~cm$^{-3}$, which is within the range 
of post-shock density derived from the H$_2$O and CO excitation analysis (Table~\ref{table:models}) for the hot gas component.
Once more, the plot highlights that the physical conditions at this shock position are different 
with respect to the other selected shock spots. 
In particular, at the L1448-R4 position the observations are consistent with a C-type shock with 
pre-shock density $n = 2 \times 10^5$~cm$^{-3}$ and velocity larger than 20~km~s$^{-1}$, 
in contrast with the predictions from C-type shock models for the [O{\sc i}] emission at 
the same shock position. 
A lower compression factor with respect to L1448-B2, in the range $2.5-25$, is suggested at L1448-R4 
from the comparison between the derived pre-shock densities and the 
post-shock densities obtained from the excitation analysis. This is consistent 
with the proposed scenario in which CO and H$_2$O emissions are produced in a C-type shock.

In conclusion (see Table~\ref{table:shock} for a summary), our analysis suggests that 
at the L1448-B2 shock position the H$_2$O and CO emissions are produced in a low-velocity 
non-dissociative J-type shock along the outflow cavity walls, 
whereas the [O{\sc i}] and maybe the OH emission originate in a fast dissociative shock.
The bright and velocity-shifted [O{\sc i}] emission at 63~$\mu$m, along with 
the detection of the [Fe{\sc ii}] line and the high [O{\sc i}]63/145$\mu$m line ratio ($\sim 20$),
supports the presence of fast dissociative
shocks related to the presence of an embedded atomic jet near the protostar
\citep[e.g.][]{hollenbach1989,flower2010}.

At the other shock positions, 
we can conclude that the excited H$_2$O and high-$J$ CO emissions are produced in a C-type shock 
with velocity greater than 20~km~s$^{-1}$, whereas a partially dissociative J-type shock 
is needed to explain the [O{\sc i}] emission. 
As discussed in Sect.~\ref{subsect:other_spots}, 
the H$_2$O abundance of the hot gas at these positions appears to be higher 
than at L1448-B2 by a factor of $\sim 4$. 
This is consistent with L1448-B2 being the signpost of a J-type shock, in which the predicted H$_2$O 
abundance is of the order of $2 \times 10^{-5}$ \citep[see][]{flower2010}.

Finally, our results are also consistent with previous HIFI observations by \citet{santangelo2012},
showing that the H$_2$O line ratios at L1448-B2 are consistent with a non-dissociative J-type shock, with 
pre-shock density $n = 2$~10$^4$~cm$^{-3}$. On the other hand, the authors found that 
in L1448-R4 the shock conditions of the low-velocity component, which dominates the emission 
in the relatively higher excitation lines, are more degenerate and a C-type shock origin could not be ruled out. 
The same degeneracy has been inferred for the two positions along the L1157 outflow by \citet{vasta2012}, 
thus consistent with a possible C-type shock origin for the H$_2$O emission.

We point out, however, that any comparison with available shock models can only be roughly indicative 
of the real physical situation occurring in the investigated shock events. 
In particular, geometrical complexity as well as chemical effects induced by diffuse UV fields (both from the
star and from associated fast shocks) would need to be properly included in a more realistic model.

%
%______________________________________________________________

\section{Conclusions}
\label{sect:conclusions}

\emph{Herschel}-PACS observations of H$_2$O, high-$J$ CO, [O{\sc i}], and OH toward two selected positions along 
the bright outflows L1448 and L1157 have been presented, as part of the WISH key program.
The main conclusions of this work are the following:
\begin{enumerate}
\item Consistent with other studies, at all selected shock positions we find a close spatial association, 
at the angular resolution of our PACS observations,
between H$_2$O emission and high-$J$ CO emission, whereas the low-$J$ CO emission seems to 
trace a different gas component, not directly associated with shocked gas.
A spatial association is also found between H$_2$O emission and mid-IR H$_2$ emission at all selected 
positions.
Moreover, no shift is found at this angular resolution between H$_2$O, [O{\sc i}], and [Fe{\sc ii}] emission, 
although the H$_2$O emission appears to be more extended than [O{\sc i}] and [Fe{\sc ii}].
\item The excitation conditions at the L1448-B2 shock position close to the driving outflow source indicate 
a two-component model to reproduce the H$_2$O and CO emission. 
In particular, an extended warm component with temperature $T \sim 450$~K 
and density $n$(H$_2$)$=10^6$~cm$^{-3}$ is traced by the bulk of the HIFI H$_2$O emission ($E_{\rm u}= 53-137$~K) 
and by the PACS CO emission up to $J = 22-21$; furthermore, 
a compact hot component with $T=1100$~K and density $n$(H$_2$)$=(0.5-5) \times 10^6$~cm$^{-3}$
is traced by the bulk of the PACS higher-excitation H$_2$O emission ($E_{\rm u} > 190$~K) 
and by the PACS higher-$J$ CO emission.
A similar stratification of gas components at different temperatures has been found for the Spitzer H$_2$ gas.
\item Among the selected positions L1448-B2 is found to be peculiar, possibly because of its proximity 
to the central driving source of the L1448 outflow. 
In particular, a non-dissociative J-type shock at the point of impact of the jet on the cloud
seems to be responsible for the H$_2$O and CO hot gas component
at this position, whereas a C-type shock is needed to explain the origin of the hot component 
at the other selected positions.
On the other hand, the observations suggest a dissociative J-type shock at L1448-B2, related to the presence 
of an embedded atomic jet, to explain the observed OH and [Fe{\sc ii}] emission and the 
bright and velocity-shifted [O{\sc i}] emission. A J-type shock that is at least partially dissociative 
is needed to explain the [O{\sc i}] emission at the other selected positions as well.
\item From the comparison between H$_2$O and H$_2$, at L1448-B2 we obtain a H$_2$O abundance of 
$(3-4) \times 10^{-6}$ for the 
warm component and of $(0.3-1.3) \times 10^{-5}$ for the hot component.
At the other examined shock positions the H$_2$O abundance of the hot component
appears to be higher by a factor of $\sim 4$, reflecting
evolutionary effects on the timescales of the outflow propagation.
The indication that the H$_2$O abundance may be higher 
in the hotter gas in some shock positions is in line with ISO data 
by other authors \citep[e.g.][]{giannini2001}.
This result is also consistent with L1448-B2 being closer to the driving outflow source 
than the other selected positions. This makes it more affected by the strong FUV radiation field 
coming from the nearby protostar that may photodissociate H$_2$O 
in the post-shock gas and thus decrease the H$_2$O abundance.
An estimate of the CO abundance was also derived at L1448-B2 and 
is of the order of $(3-4) \times 10^{-5}$ for the warm component, whereas it is $(1-2) \times 10^{-5}$ 
for the hot component. 
\item These results, along with the spatial extent inferred for the different gas components, lead us to the 
conclusion that the two gas components represent a gas stratification in the post-shock region.
In particular, the 
extended and low-abundance warm component traces the post-shocked gas that has already cooled down 
to a few hundred Kelvin, 
whereas the compact and possibly more abundant hot component is associated with the gas 
that is currently undergoing a shock episode, being compressed and heated to a thousand Kelvin.
This hot gas component is thus possibly affected by evolutionary effects 
on the timescales of the outflow propagation, which explains the variations of H$_2$O abundance 
we observed at the different positions along the outflows.
\end{enumerate}

\begin{acknowledgements}
WISH activities at Osservatorio Astronomico di Roma are supported by the ASI project 01/005/11/0.
G.S. and B.N. also acknowledge financial contribution from the agreement
ASI-INAF I/009/10/0.
Astrochemistry in Leiden is supported by NOVA, by a Spinoza grant and
grant 614.001.008 from NWO, and by EU FP7 grant 238258.
HIFI has been designed and built by a consortium of 
institutes and university departments from across Europe, Canada and the 
United States under the leadership of SRON Netherlands Institute for Space
Research, Groningen, The Netherlands and with major contributions from 
Germany, France and the US. Consortium members are: Canada: CSA, 
U.Waterloo; France: CESR, LAB, LERMA, IRAM; Germany: KOSMA, 
MPIfR, MPS; Ireland, NUI Maynooth; Italy: ASI, IFSI-INAF, Osservatorio 
Astrofisico di Arcetri- INAF; Netherlands: SRON, TUD; Poland: CAMK, CBK; 
Spain: Observatorio Astron{\'o}mico Nacional (IGN), Centro de Astrobiolog{\'i}a 
(CSIC-INTA). Sweden: Chalmers University of Technology - MC2, RSS $\&$ 
GARD; Onsala Space Observatory; Swedish National Space Board, Stockholm 
University - Stockholm Observatory; Switzerland: ETH Zurich, FHNW; USA: 
Caltech, JPL, NHSC.
\end{acknowledgements}

\Online

\begin{appendix} %First online appendix

\section{PACS maps}
\label{appendix:pacs_maps}

\begin{figure*}%[h!]
   \centering
   \includegraphics[angle=-90,width=0.95\textwidth]{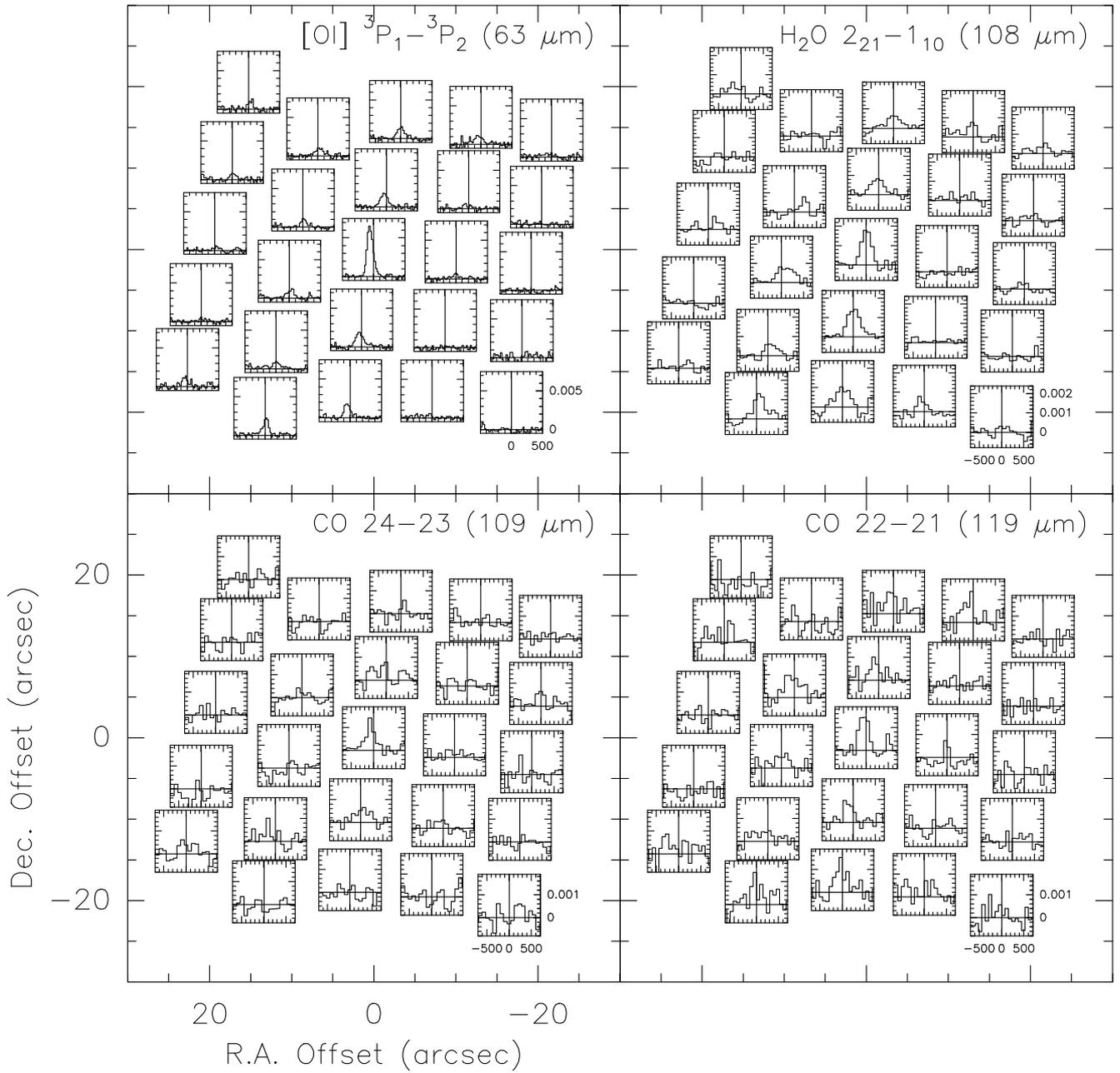}
   \caption{PACS spectra of the detected transitions at the L1448-B2 position.
The centre of each spaxel box corresponds to its offset position with respect to the coordinates 
of the central shock position.
The velocity (km~s$^{-1}$) and intensity scale (Kelvin) are indicated for one spaxel box and refer to all spectra 
of the relative transition.
The labels in the top-right corner of every box indicate the relative transition.
}
   \label{fig:PACSl1448B2}
   \end{figure*}

\begin{figure*}%[h!]
   \centering
   \includegraphics[angle=-90,width=0.95\textwidth]{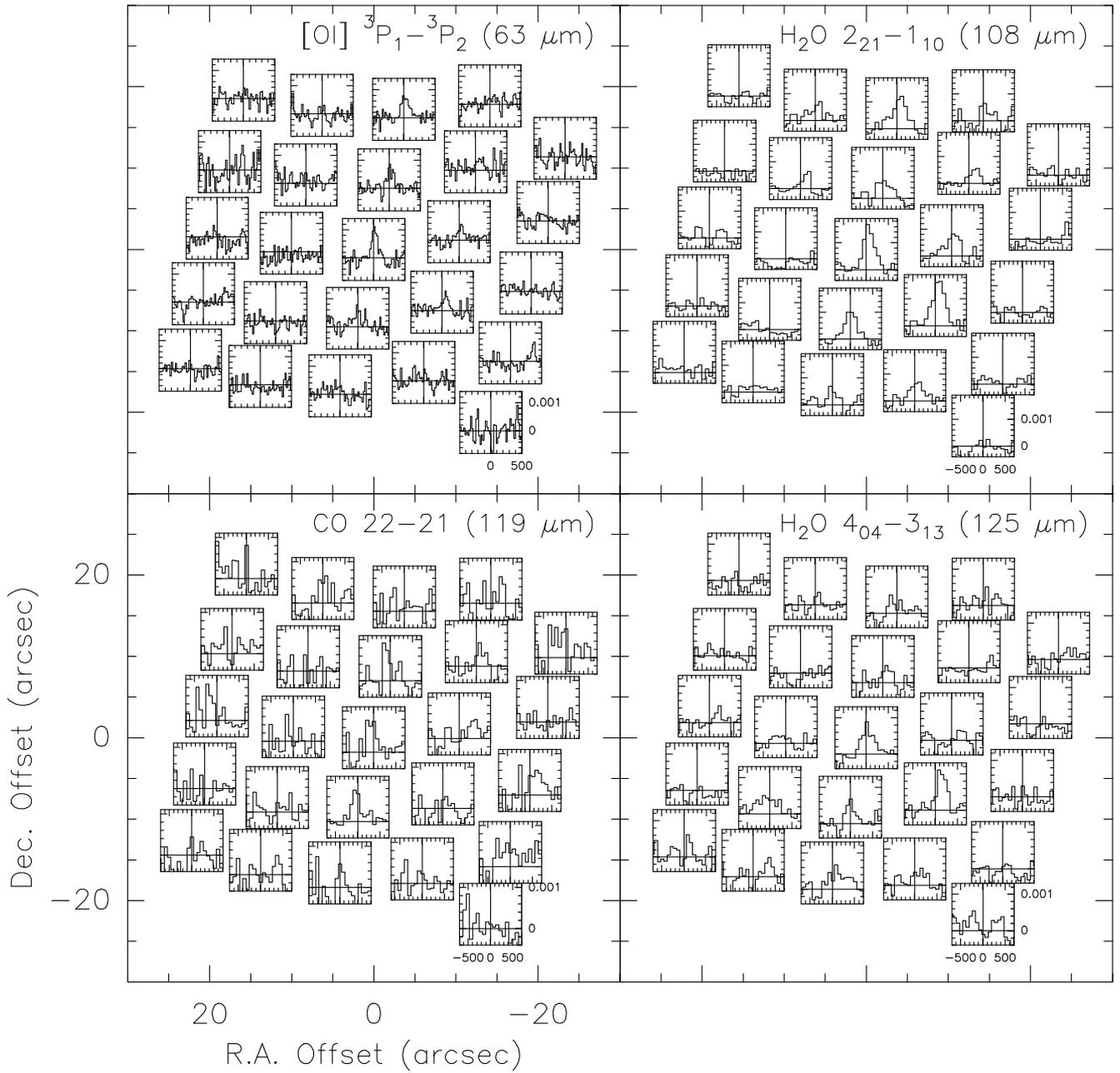}
   \caption{Same as Fig.~\ref{fig:PACSl1448B2} for the L1448-R4 position.
}
   \label{fig:PACSl1448R4}
   \end{figure*}

\begin{figure*}%[h!]
   \centering
   \includegraphics[angle=-90,width=0.95\textwidth]{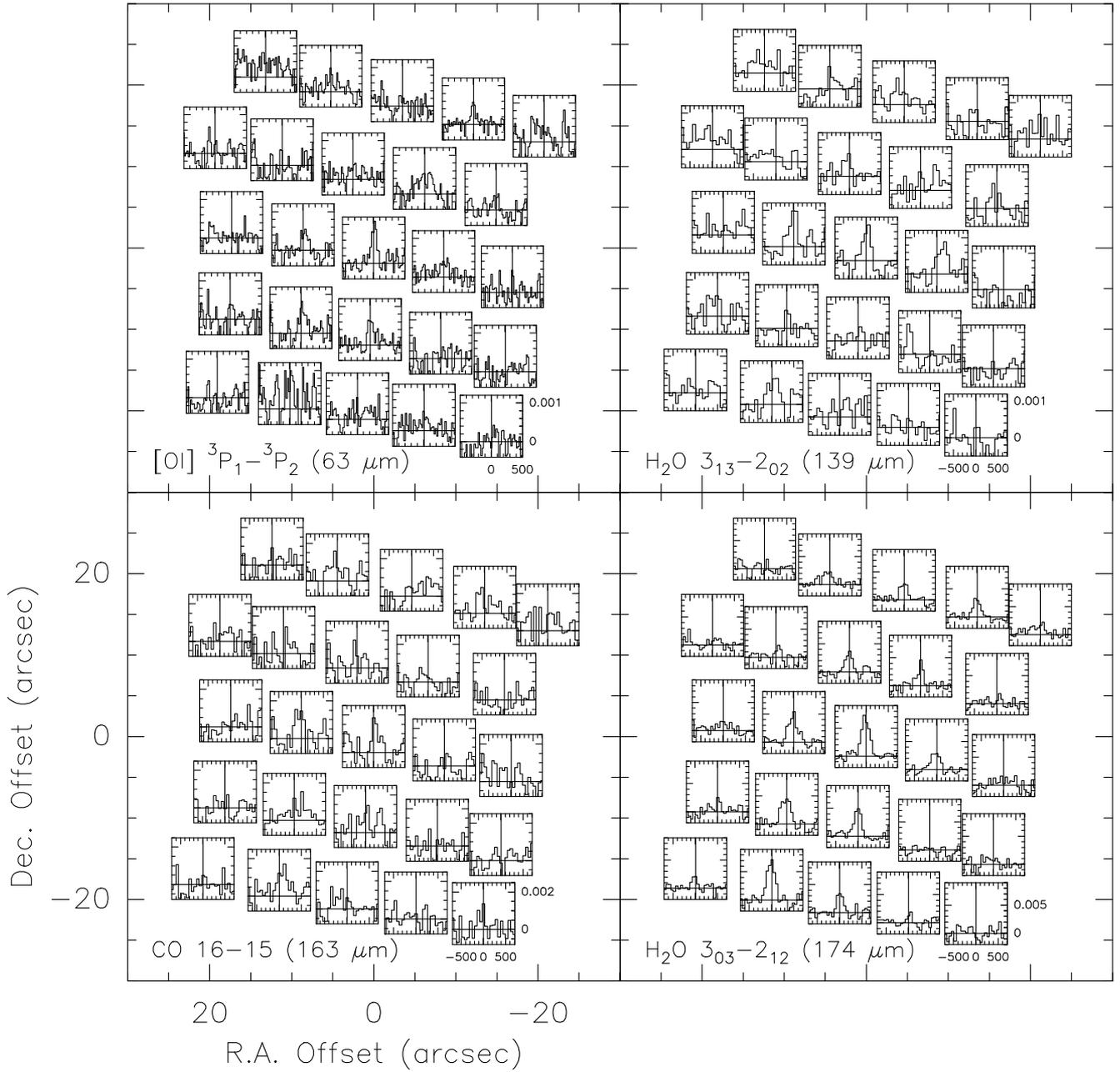}
   \caption{Same as Fig.~\ref{fig:PACSl1448B2} for the L1157-B2 position.
}
   \label{fig:PACSl1157B2}
   \end{figure*}

\begin{figure*}%[h!]
   \centering
   \includegraphics[angle=-90,width=0.95\textwidth]{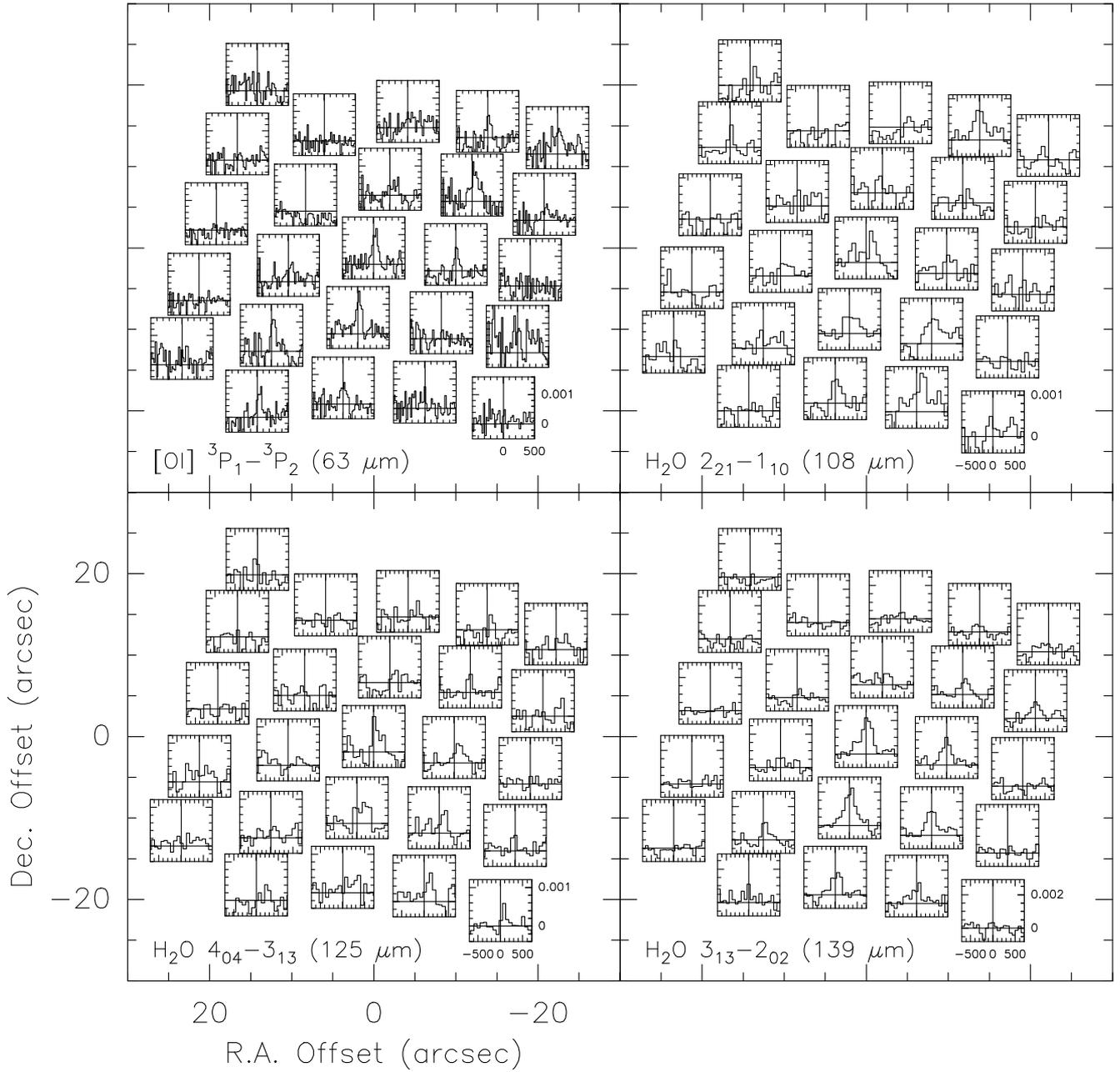}
   \caption{Same as Fig.~\ref{fig:PACSl1448B2} for the L1157-R position.
}
   \label{fig:PACSl1157R}
   \end{figure*}

The PACS maps of all the lines observed at the 
four selected shock positions (B2 and R4 along L1448 and B2 and R along L1157) are 
presented in this section (see also Table~\ref{table:intensities}) . 
In particular, Figs.~\ref{fig:PACSl1448B2} and \ref{fig:PACSl1448R4} are relative 
to the L1448 outflow (B2 and L1157 positions, respectively), 
whereas Figs.~\ref{fig:PACSl1157B2} and \ref{fig:PACSl1157R} concern the L1157 outflow
(B2 and R positions, respectively).

   \begin{figure*}%[h!]
   \centering
   \includegraphics[angle=-90,width=0.95\textwidth]{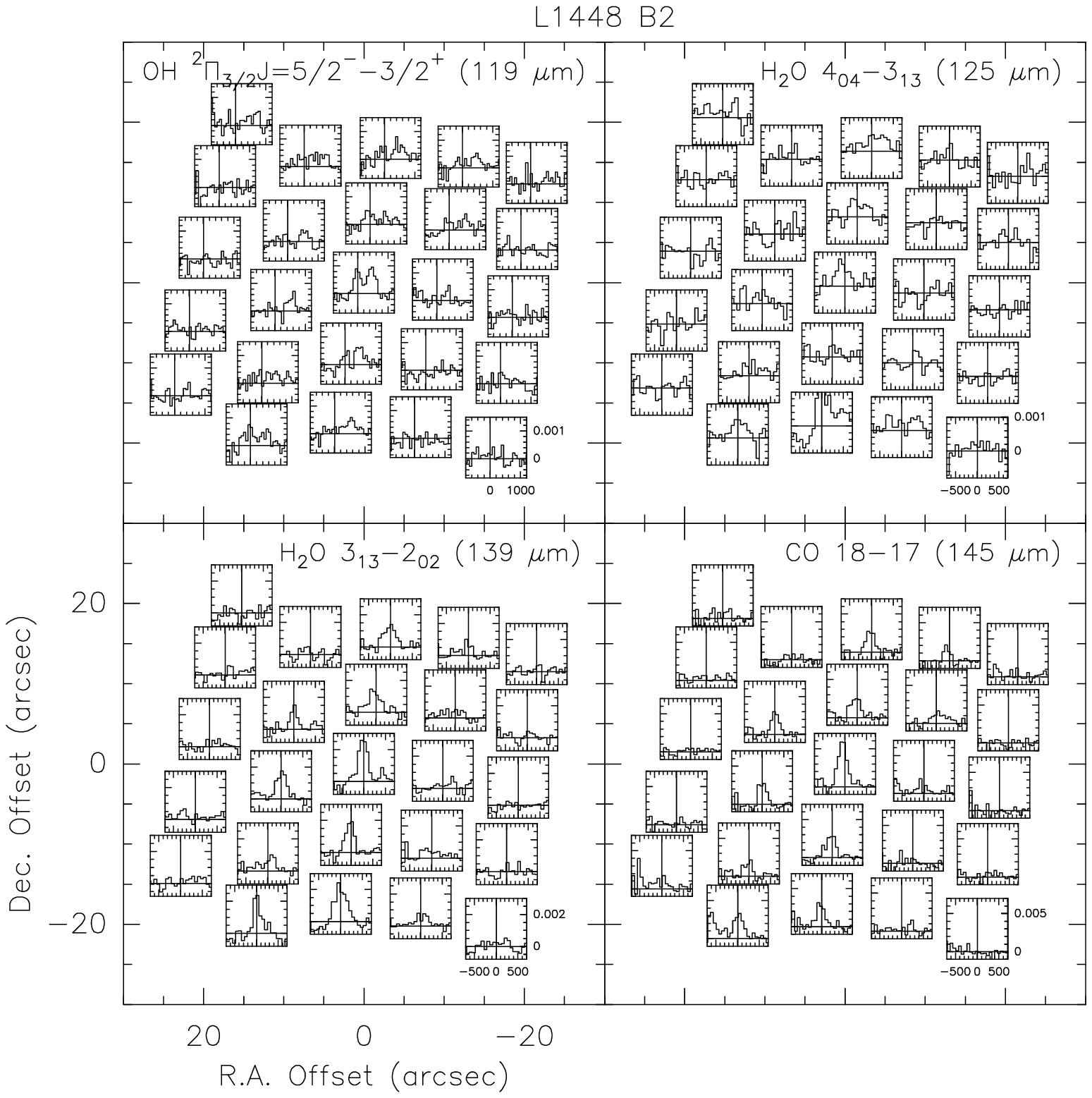}
   \begin{center}
   Fig.~\ref{fig:PACSl1448B2} -- Continued.
   \end{center}
   \end{figure*}
   \begin{figure*}%[h!]
   \centering
   \includegraphics[angle=-90,width=0.95\textwidth]{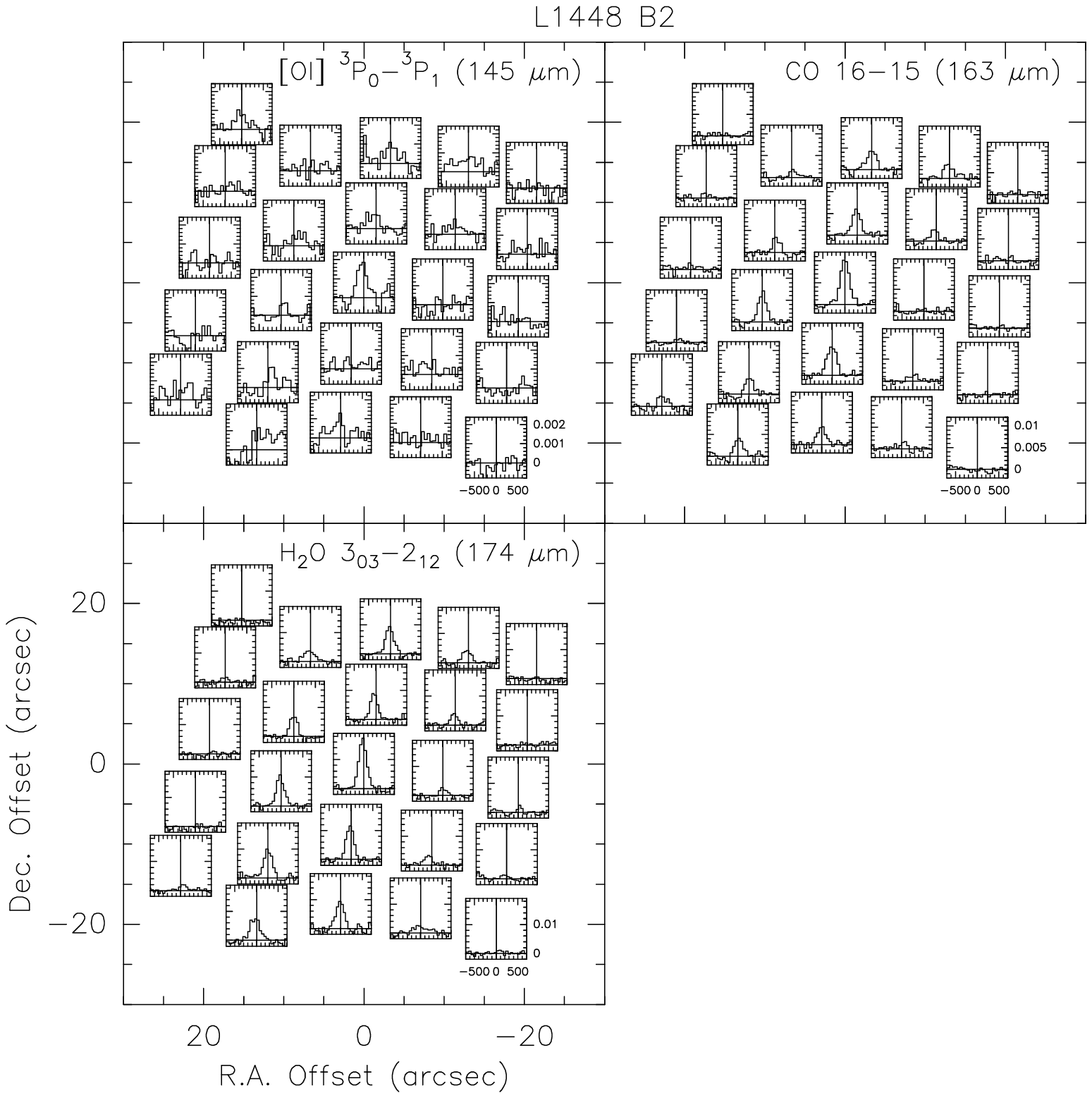}
   \begin{center}
   Fig.~\ref{fig:PACSl1448B2} -- Continued.
   \end{center}
   \end{figure*}

   \begin{figure*}%[h!]
   \centering
   \includegraphics[angle=-90,width=0.95\textwidth]{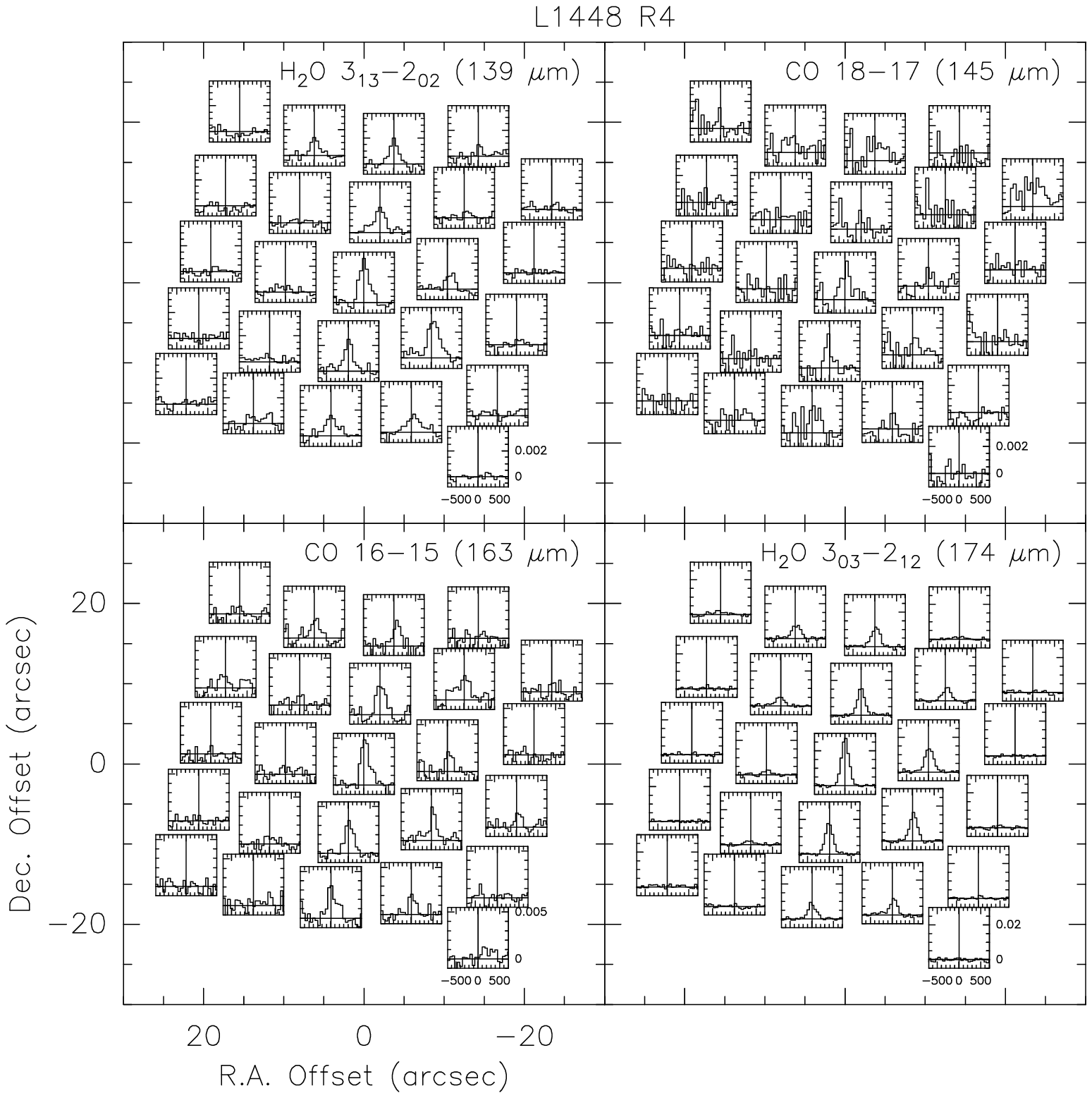}
   \begin{center}
   Fig.~\ref{fig:PACSl1448R4} -- Continued.
   \end{center}
   \end{figure*}

   \begin{figure*}%[h!]
   \centering
   \includegraphics[angle=-90,width=0.95\textwidth]{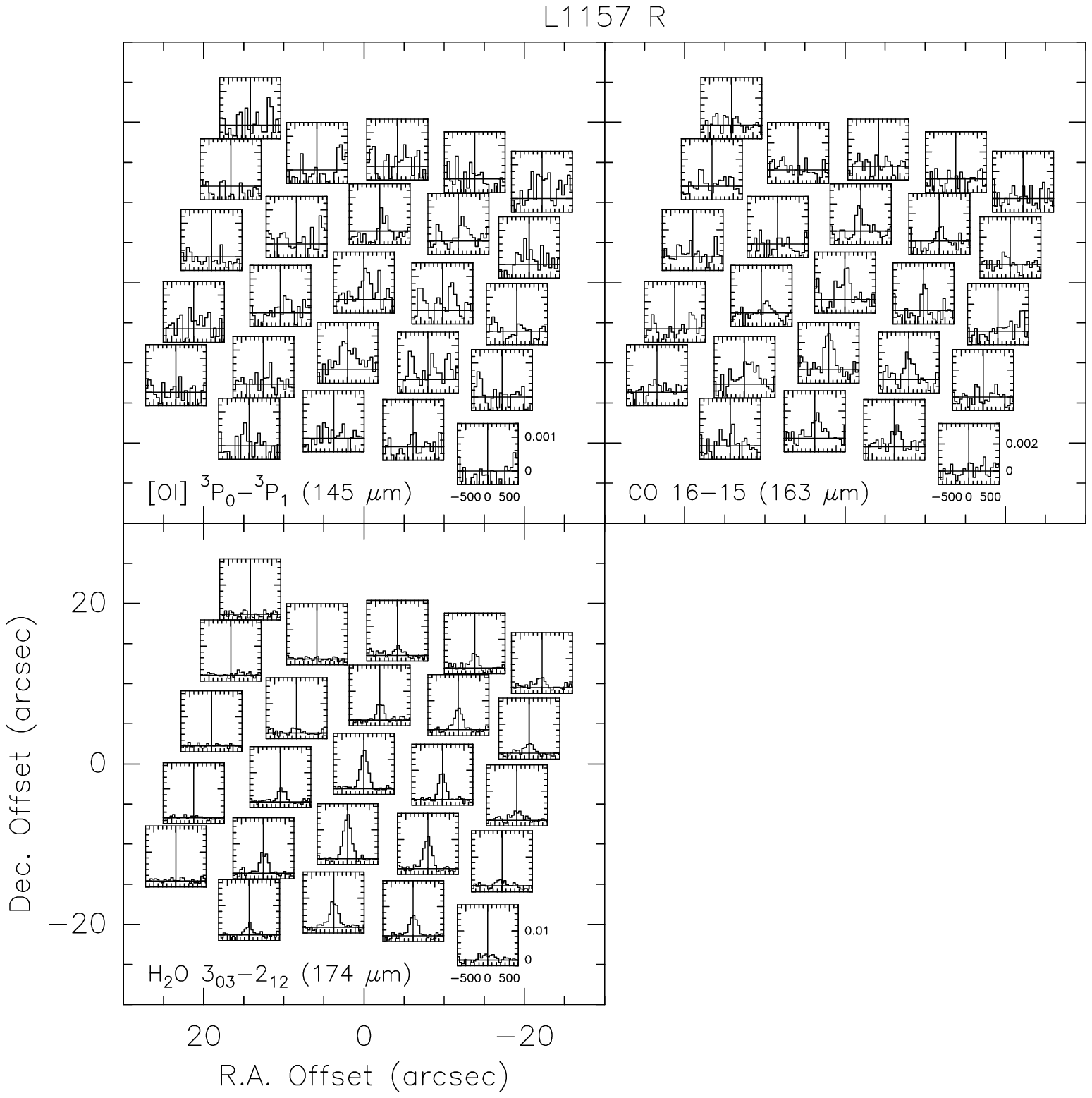}
   \begin{center}
   Fig.~\ref{fig:PACSl1157R} -- Continued.
   \end{center}
   \end{figure*}

\end{appendix}

\end{document}